\documentclass[10pt]{amsart}
\usepackage{latexsym,amsfonts,amsmath,amssymb}
\usepackage[all]{xy}
\usepackage{a4wide}

\newcommand{\IR}{{\mathbb{R}}}
\newcommand{\ID}{{\mathbb{D}}}
\newcommand{\ZZ}{{\mathbb{Z}}}
\newcommand{\IC}{{\mathbb{C}}}

\newcommand{\C}{{\mathcal{C}}}

\newcommand{\RR}{{{}_{R}}}

\newcommand{\al}{{\alpha}}
\newcommand{\ze}{{\zeta}}
\newcommand{\la}{{\lambda}}
\newcommand{\na}{{\nabla}}

\newcommand{\g}{{\mathfrak{g}}}
\newcommand{\A}{{\mathfrak{a}}}
\newcommand{\U}{{\mathbb{U}}}
\newcommand{\LL}{{\mathfrak{l}}}
\newcommand{\LLL}{{\mathbb{L}}}

\newcommand{\mut}{{\tilde\mu}}
\newcommand{\Ft}{{\tilde{F}}}

\newcommand{\spec}{{\rm{spec}}}
\newcommand{\Tr}{{\rm{Tr}}}

\newcounter{smalllist}

\newtheorem{theorem}{Theorem}

\newtheorem{lemma}{Lemma}[section]
\newtheorem{prop}[lemma]{Proposition}
\newtheorem{coro}[lemma]{Corollary}
\theoremstyle{definition}

\newtheorem{remark}[lemma]{Remark}

\let\Re=\undefined\DeclareMathOperator{\Re}{Re}
\let\Im=\undefined\DeclareMathOperator{\Im}{Im}
\DeclareMathOperator{\diag}{diag} 

\let\llldots=\ldots
\def\ldots{\llldots{}}

\numberwithin{equation}{section}

\begin{document}

\title[Multi-Hamiltonian structure for AL]{Multi-Hamiltonian structure for the finite defocusing Ablowitz-Ladik equation}
\author[M.~Gekhtman and I.~Nenciu]{Michael Gekhtman and Irina Nenciu}
\address{Michael Gekhtman\\
         Department of Mathematics\\
         University of Notre Dame\\
         Notre Dame, IN 46556}
\email{Michael.Gekhtman.1@nd.edu}
\address{Irina Nenciu\\
         Courant Institute\\
         251 Mercer Street\\
         New York, NY 10012 \textit{and} Institute of Mathematics ``Simion Stoilow''
     of the Romanian Academy\\ 21, Calea Grivi\c tei\\010702-Bucharest, Sector 1\\Romania}
\email{nenciu@cims.nyu.edu}

\begin{abstract}
We study the Poisson structure associated to the defocusing Ablowitz-Ladik equation from
a functional-analytical point of view, by reexpressing the Poisson bracket in terms
of the associated Carath\'eodory function. Using this expression, we are able to
introduce a family of compatible Poisson brackets which form a multi-Hamiltonian
structure for the Ablowitz-Ladik equation. Furthermore, we show using some of these
new Poisson brackets that the Geronimus relations between orthogonal polynomials
on the unit circle and those on the interval define an algebraic and symplectic mapping
between the Ablowitz-Ladik and Toda hierarchies.
\end{abstract}
\thanks{M. G. is partially supported by the NSF grant $\#$ 0400484. I.N.'s research was partly supported
by NSF grant DMS-0111298, and was done while she was a member of the Institute for Advances Study, Princeton. The authors wish to thank Percy Deift, Peter Miller and Barry Simon for useful discussions.}
\maketitle

\tableofcontents

\section{Introduction}\label{S:I}

It has been well-known since the work of Flaschka~\cite{Fla1},
\cite{Fla} that the celebrated Toda lattice can be represented as an
isospectral evolution equation for Jacobi matrices, by which we mean
symmetric, tridiagonal matrices $J$, with positive off-diagonals.
This not only allowed Flaschka and H\'enon to prove the complete
integrability of the Toda equation, but also opened the door for the
study of related Lie algebraic and geometric structures for
generalized Toda flows with applications ranging from numerical
analysis to quantum cohomology. We will not attempt to properly cite
vast literature on the Toda lattice, instead referring the reader to
the bibliography in \cite{IntSysII} or in the recent survey
\cite{BG}. We will mention, however, two seminal works that
introduced two aspects of the theory crucial to our exposition:
Moser's paper \cite{Moser} that explained the role played by
spectral data in both explicitly solving and establishing complete
integrability of the Toda lattice  and Kostant's comprehensive
treatment of the generalized Toda lattice as a restriction of the
larger Hamiltonian system to a minimal irreducible coadjoint orbit
of the Borel subgroup (the manifold of Jacobi matrices with fixed trace in the case of
$gl(n)$) equipped with the Lie-Poisson structure. Combination of
these two approaches later allowed Deift \textit{et al.} to
establish complete integrability of full Toda flows
\cite{DLNT_CPAM86}. More recently, it was shown in \cite{FG2}, that
the spectral data can be used to define in a natural way a
multi-Hamiltonian structure for a family of Toda-like systems
associated with all minimal irreducible coadjoint orbits in
$gl(n)$.

In this paper we concentrate on the study of another integrable system, the defocusing Ablowitz-Ladik (AL) hierarchy,
through its connection to a unitary analogue of Jacobi matrices, the so-called CMV matrices. The defocusing AL equation
was defined in 1975--76 by Ablowitz and Ladik \cite{AL1,AL2} as a
space-discretization of the cubic nonlinear Schr\"odinger equation.
It reads:
\begin{equation}\label{ALE1}
-i\dot\beta_k=\rho_k^2 (\beta_{k+1}+\beta_{k-1})-2\beta_k,
\end{equation}
where $\beta=\{\beta_k\}\subset\mathbb{D}$ is a sequence of
complex numbers inside the unit disk and
$\rho_k^2=1-|\beta_k|^2$.
The analogy with the continuous NLS becomes transparent if we
rewrite \eqref{ALE1} as\footnote{Here, and throughout the paper, $\dot f$ will denote the time
derivative of the function $f$.}
$$
-i\dot\beta_k=\beta_{k+1}-2\beta_k+\beta_{k-1}
-|\beta_k|^2(\beta_{k+1}+\beta_{k-1}).
$$
 A simple change of variables,
$\al_k=e^{2it}\beta_k$ for all $k$, transforms \eqref{ALE1} into
\begin{equation}\label{ALE2}
-i\dot\al_k=\rho_k^2\bigl(\al_{k+1}+\al_{k-1}\bigr),
\end{equation}
where $\rho_k=\sqrt{1-|\al_k|^2}$. This is the equation we will refer to as Ablowitz-Ladik (or AL). In this paper we
focus on the finite case, by which we mean the case in which $\al_{-1}=-1$ and,
for some fixed $n\geq 1$, $\al_{n-1}\in S^1=\{z\in\IC\,|\, |z|=1\}$. These are
Dirichlet boundary conditions, and one can easily see from \eqref{ALE2} that
the evolution for $\al_0,\dots,\al_{n-2}$ decouples from the evolution for the other $\al$'s.
Consider the following Poisson bracket on the space of $(\al_0,\dots,\al_{n-2},\al_{n-1})\in\ID^{n-1}\times S^1$:
\begin{equation}\label{E:PbDefn}
\{f,g\}=i\sum_{k=0}^{N-2} \rho_k^2 \left[\frac{\partial f}{\partial \bar\al_k}\frac{\partial g}{\partial \al_k}
-\frac{\partial f}{\partial \al_k}\frac{\partial g}{\partial \bar\al_k}\right],
\end{equation}
where $\rho_k=\sqrt{1-|\al_k|^2}$ and, for a complex variable $\al=u+iv$, $u,v\in\IR$, the partial derivatives are defined
as usual by
\begin{equation*}
\frac{\partial}{\partial\al}=\frac12\left[\frac{\partial}{\partial u}-i\frac{\partial}{\partial v}\right],
\qquad
\frac{\partial}{\partial\bar\al}=\frac12\left[\frac{\partial}{\partial u}+i\frac{\partial}{\partial v}\right].
\end{equation*}
In this Poisson structure, the AL equation \eqref{ALE2} becomes
completely integrable. Moreover, it can be re-written in the Lax
form, with the Hamiltonian $\Re\Tr \C$, where the Lax operator $\C$
is the CMV matrix associated with the coefficients
$\al_0,\dots,\al_{n-2},\al_{n-1}$ (for the background, see
Subsection~\ref{Ss:2.1}). In fact, one can define a whole hierarchy
of evolution equations, that we will call the AL hierarchy, by
considering the Hamiltonians given by the real and imaginary parts
of $K_k=\frac1k \Tr(\C^k)$ for $k\geq1$. In terms of Lax pairs, the hierarchy is given by the
following evolutions equations (see \cite{N}):
\begin{equation}\label{E:ALhreal}
\{\C,2\Re (K_k)\}=[\C,i(\C^k)_+ +i((\C^k)_+)^*]
\end{equation}
and
\begin{equation}\label{E:ALhimag}
\{\C,2\Im (K_k)\}=[\C,(\C^k)_+ -((\C^k)_+)^*]
\end{equation}
for all $k\geq1$, where for a matrix $X$, we have
\begin{equation*}
(X_+)_{jk}=\left\{%
\begin{array}{ll}
    X_{jk}, & \quad\hbox{if}\,\,j<k;\\
    \tfrac12X_{jj}, & \quad\hbox{if}\,\,j=k;\\
    0, & \quad\hbox{if}\,\,j>k.\\
\end{array}%
\right.
\end{equation*}

One of the central ingredients in the study of the AL hierarchy is, similarly to
the Toda case, rewriting the Poisson bracket \eqref{E:PbDefn} as the restriction to the manifold
of CMV matrices of the Gelfand-Dikij bracket on the associative  algebra $M_n(\IC)$  of $n\times n$ matrices (for a short background, see Subsection~\ref{Ss:2.2}). This was done
independently by L. C. Li~\cite{L}, and R. Killip and I. Nenciu~\cite{KN}, and allowed
Killip and Nenciu to solve the system and find the long-time asymptotics of the $\al$s, and of
certain associated spectral quantities.  Inverse spectral problem  for semi-infinite CMV matrices
was recently utilized in this context in \cite{Gol}. Note also that an alternative Lax representation  was used
to linearize  finite and semi-infinite AL flows in \cite{G}, while the approach based on continued fractions
was suggested in \cite{Common}.

If  $d\mu$ is the
measure on the unit circle associated to the Verblunsky coefficients $(\al_0,\dots,\al_{n-2},\al_{n-1})\in\ID^{n-1}\times S^1$,
then it is known (see Subsection~\ref{Ss:2.1}) that $d\mu=\sum_{j=1}^n \mu_j\delta_{z_j}$, with $\mu_j\in(0,1)$,
$\mu_1+\cdots+\mu_n=1$, and $z_j=e^{i\theta_j}\in S^1$ for all $1\leq j\leq n$. The function
$$
\prod_{j=1}^n z_j=\det(\C)=(-1)^{n-1} \bar\al_{n-1}
$$
is a Casimir, and the manifold of CMV matrices with fixed determinant forms a symplectic leaf on which,
for $1\leq j,k\leq n-1$\footnote{Since
the $z_j$s are always distinct, any choice of labeling for $z_1,...,z_n$ is locally well-defined and
leads to these formulae.}, we have
\begin{equation}\label{E:6}
\{\theta_j,\theta_k\}=0,\quad \{\theta_j,\tfrac12\log[\mu_k/\mu_n]\}=\delta_{jk},
\end{equation}
and
\begin{equation}\label{E:OhDear}
\bigl\{ \log[\mu_j/\mu_n] , \log[\mu_k/\mu_n] \bigr\} = 2\cot\bigl(\tfrac{\theta_j-\theta_k}2\bigr)
+ 2\cot\bigl(\tfrac{\theta_k-\theta_n}2\bigr)
+ 2\cot\bigl(\tfrac{\theta_n-\theta_j}2\bigr).
\end{equation}
These results are obtained by regarding the CMV matrices, or, equivalently, the associated spectral measures,
as the central objects. In this paper, we adopt a slightly different point of view, and focus on
the associated {\it Carath\'eodory function},
$$
F(z)=\left(\frac{\C+z}{\C-z}\right)_{11}.
$$
This is the analogue in the unitary case of the better known \textit{Weyl function} (or $m$-function) associated to a Jacobi matrix,
$$
m(\lambda)=\left(\frac{1}{J-\lambda}\right)_{11}.
$$
In \cite{FG2}, Faybusovich and Gekhtman adopted this point of view
for the Toda lattice and computed Poisson brackets induced by the
Lie-Poisson structure for $m(\lambda)$ and $m(\xi)$ at any two
distinct points $\lambda$ and $\xi$. The resulting Poisson structure
on Weyl functions was then shown to be a part  of a  family of
compatible Poisson brackets which constitutes a multi-Hamiltonian
structure for the Toda lattice. In this paper, we follow the same
road in the Ablowitz-Ladik case.

The paper is organized as follows. In Section~\ref{S:2} we give some
background information on the theory of orthogonal polynomials on
the real line and unit circle, and on classical $R$-matrices.
Section~\ref{S:3} contains the first important results,
Theorem~\ref{T:CSbracket} and Corollary~\ref{C:3.1}, which gives the
formula for the Poisson bracket of the Carath\'eodory function at
two distinct points in the complex plane. In particular, this
represents a more direct proof of some of the results in \cite{KN}.
Here we should also mention that in a recent paper \cite{CanSim}, M.
Cantero and B. Simon study Poisson brackets for orthogonal
polynomials both on the real line and on the unit circle induced by
{\em standard} Poisson structures for the Toda and Ablowitz-Ladik
hierarchies, respectively. An essential part of their analysis are
the formulae for the Poisson brackets of Weyl and Carath\'eodory
functions, for which they give new proofs by induction, using purely
orthogonal polynomial methods. Another related recent paper is
\cite{Tsi}, which also introduces a family of Poisson brackets
compatible with the Sklyanin  bracket associated with the standard
$2\times2$ {\em rational} solution of the Classical Yang-Baxter
equation. These Poisson brackets are defined on monodromy matrices
associated with $2\times 2$ spectral parameter depending  Lax
representation for a family of integrable systems that includes both
open and periodic Toda lattices.

The formula \eqref{E:Cbrac} for the Poisson bracket mentioned above
allows us to extend the Poisson structure to the space of finite,
but unnormalized measures on the circle, and obtain the canonical
coordinates for both the extended and the usual Poisson structures;
this is achieved in Section~\ref{S:ExtendedPB}. Finally, we define
the family of compatible Poisson structures in
Section~\ref{S:CompatiblePB}, and show its connection to the
defocusing Ablowitz-Ladik equation. In the last section,
Section~\ref{S:6}, we go back to the Toda lattice and show that,
loosely speaking, `half' of the Ablowitz-Ladik hierarchy (also known
as the Schur flows) is mapped symplectically onto the Toda lattice hierarchy via
the well-known Geronimus relations.

\section{Background}\label{S:2}

\subsection{Orthogonal polynomials}\label{Ss:2.1}

As CMV matrices arose in the study of orthogonal polynomials, it is
natural that we begin there. We will first describe the relation of
orthogonal polynomials to Jacobi matrices and then explain the
connection to CMV matrices.

Given a probability measure $d\nu$ supported on a finite subset of
$\IR$, say of cardinality $n$, we can apply the Gram--Schmidt
procedure to $\{1,x,x^2,\ldots,x^{n-1}\}$ and so obtain an
orthonormal basis for $L^2(d\nu)$ consisting of polynomials,
$\{p_j(x):j=0,\ldots,{n-1}\}$, with positive leading coefficient. In
this basis, the linear transformation $f(x)\mapsto xf(x)$ is
represented by a Jacobi matrix,
\begin{equation}\label{Jmat}
J=\begin{bmatrix}
 b_1  &  a_1 &        & \\
 a_1  &  b_2 & \ddots & \\
      &\ddots& \ddots & a_{n-1}\\
      &      & a_{n-1}&  b_n
\end{bmatrix}
\end{equation}
with $a_j>0$, $b_j\in\IR$.  An equivalent statement is that the
orthonormal polynomials obey a three-term recurrence:
$$
xp_{j}(x) = a_{j} p_{j+1}(x) + b_{j}p_j(x) + a_{j-1} p_{j-1}(x)
$$
where $a_{-1}=0$ and $p_{n}\equiv 0$.  A third equivalent statement
is the following:  $\lambda$ is an eigenvalue of $J$ if and only if
$\lambda\in\text{supp}(d\nu)$; moreover, the corresponding eigenvector is
$[p_{0}(\lambda),p_{1}(\lambda),\ldots,p_{n-1}(\lambda)]^{T}$.

We have just shown how measures on $\IR$ lead to Jacobi matrices;
in fact, there is a one-to-one correspondence between them.  Given a
Jacobi matrix, $J$, let $d\nu$ be the spectral measure associated to
$J$ and the vector $e_1=[1,0,\ldots,0]^{T}$.  Then $J$ represents
$x\mapsto xf(x)$ in the basis of orthonormal polynomials associated
to $d\nu$.

Before explaining the origin of CMV matrices, it is necessary to
delve a little into the theory of orthogonal polynomials on the unit
circle.  For a more complete description of what follows, the reader
should turn to \cite{Simon1}.
Given a finitely-supported probability measure $d\mu$ on $S^1$, the
unit circle in $\IC$, we can construct an orthonormal system of
polynomials, $\phi_k$, by applying the Gram--Schmidt procedure to
$\{1,z,z^2,\ldots\}$.  These obey a recurrence relation; however, to
simplify the formulae, we will present the relation for the monic
orthogonal polynomials $\Phi_k(z)$:
\begin{align}
\Phi_{k+1}(z)   &= z\Phi_k(z)   - \bar\alpha_k \Phi_k^*(z).
\label{PhiRec}
\end{align}
Here $\alpha_k$ are recurrence coefficients, which are called
Verblunsky coefficients, and $\Phi_k^*$ denotes the reversed
polynomial:
\begin{equation}\label{rev}
\Phi_k(z) = \sum_{l=0}^k c_l z^l \quad \Rightarrow \quad \Phi_k^*(z)
= \sum_{l=0}^k \bar{c}_{k-l} z^l=z^k\overline{\Phi_k(1/\bar z)}.
\end{equation}
When $d\mu$ is supported at exactly $n$ points, $\alpha_k\in\ID$
for $0\leq k\leq {n-2}$ while $\alpha_{n-1}$ is a unimodular complex
number.  (Incidentally, if $d\mu$ has infinite support, then there are
infinitely many Verblunsky coefficients and all lie inside the unit disk.)
The Verblunsky coefficients completely describe the measure $d\mu$:
\begin{theorem}[Verblunsky]\label{T:Verbl}
There is a 1-to-1 correspondence between probability measures on the unit circle
supported at $n$ points and Verblunsky coefficients $(\alpha_0,\ldots,\alpha_{n-1})$
with $\alpha_{k}\in\ID$ for $0\leq k\leq n-2$ and $\alpha_{n-1}\in S^1$.
\end{theorem}

>From the discussion of Jacobi matrices, it would be natural to consider a matrix representation
of $f(z)\mapsto zf(z)$ in $L^2(d\mu)$.
Cantero, Moral, and Vel\'azquez had the simple and ingenious idea to define a basis in $L^2(d\mu)$ by applying the Gram--Schmidt
procedure to $\{1,z,z^{-1},z^2,z^{-2},\ldots\}$. The resulting
functions, $\chi_k(z)$ ($0\leq k\leq n-1$), are easily expressed in terms of the orthonormal polynomials:
\begin{equation}
\chi_k(z) = \begin{cases} z^{-k/2} \phi_k^*(z) & \text{: $k$ even} \\
        z^{-(k-1)/2} \phi_k(z) & \text{: $k$ odd.} \end{cases}
\end{equation}
In the orthonormal basis $\{\chi_k(z)\}$ of $L^2(d\mu)$, the operator $f(z)\mapsto zf(z)$ is
represented by the CMV matrix associated to the Verblunsky coefficients of the measure $d\mu$:
Given the Verblunsky coefficients $\alpha_0,\ldots,\alpha_{n-2}$ in $\ID$ and
$\alpha_{n-1}\in S^1$ associated to the measure $d\mu$, let $\rho_k=\sqrt{1-|\alpha_k|^2}$, and
define $2\times 2$ matrices
$$
\Xi_k = \begin{bmatrix} \bar\alpha_k & \rho_k \\ \rho_k & -\alpha_k
\end{bmatrix}
$$
for $0\leq k\leq n-2$, while $\Xi_{-1}=[1]$ and
$\Xi_{n-1}=[\bar\alpha_{n-1}]$ are $1 \times1$ matrices. From these,
form the $n\times n$  block-diagonal matrices
$$
\mathcal{L}=\diag\bigl(\Xi_0   ,\Xi_2,\Xi_4,\ldots\bigr)
\quad\text{and}\quad
\mathcal{M}=\diag\bigl(\Xi_{-1},\Xi_1,\Xi_3,\ldots\bigr).
$$
The \emph{CMV matrix} associated to the coefficients
$\alpha_0,\ldots,\alpha_{n-1}$ is $\C=\mathcal{LM}$.

The measure $d\mu$ can be reconstructed from $\C$ in a manner analogous to the Jacobi case:
\begin{theorem}\label{T:2b}
Let $d\mu$ be the spectral measure associated to a CMV matrix, $\C$,
and the vector $e_1$.  Then $\C$ is the CMV matrix associated to the
measure $d\mu$.
\end{theorem}

Proofs of these Theorems can be found in
\cite{CMV} or \cite{Simon1}. As explained in the Introduction, throughout the paper
we will always use implicitly the bijection between measures $d\mu=\sum \delta_{z_j}\mu_j$, CMV matrices, and the coordinates
given by the $z_j$'s and $\mu_j$'s. A very important notion, that will be heavily used in this paper, is
{\it the Carath\'eodory function} associated to a probability measure $\mu$ on the unit circle $S^1$; it
is given by (see \cite[Section~1.3]{Simon1})
\begin{equation}\label{E:4}
F(z)=\int \frac{e^{i\theta}+z}{e^{i\theta}-z}\,d\mu(\theta).
\end{equation}
In terms of the other coordinates, $F$ is given by
\begin{equation}\label{E:5}
F(z)=\left(\frac{\C+z}{\C-z}\right)_{11}=\sum_{j=1}^n \frac{z_j+z}{z_j-z}\mu_j,
\end{equation}
where $\C$ is the CMV matrix associated to the measure $\mu$.
Moreover, the Carath\'eodory function is related to {\it the Schur function} $f$ by
\begin{equation}\label{E:CaraSchur}
F(z)=\frac{1+zf(z)}{1-zf(z)}\qquad\Longleftrightarrow\qquad f(z)=\frac1z\,\frac{1-F(z)}{1+F(z)}.
\end{equation}
While the Carath\'eodory function plays a very important role throughout the theory of orthogonal polynomials
on the unit circle, it is very simple to see from the formulae above that, in the case of finite measures, $F$ encodes exactly
the same information as the measure $\mu$. This is exactly the reason why we can use the Carath\'eodory function
in a functional analytic approach to describe all the Poisson structures that we will introduce.

\subsection{Integrable systems and classical R-matrices}\label{Ss:2.2}

The manifold of Jacobi matrices with fixed trace forms a co-adjoint
orbit of the group of invertible upper triangular matrices, if one views the space of symmetric matrices as a dual space to the algebra of upper triangular matrices. The Lie-Poisson structure on this dual thus induces  a symplectic structure on the Jacobi orbit. More generally, the space of 3-diagonal (not necessarily symmetric) matrices form a Poisson submanifold in $gl(n;\IR)$ with respect to the
 Lie-Poisson bracket associated to a particular Lie algebra structure on the
$n\times n$ matrices (although not the one defined via the usual matrix commutator).
These matters are
described in detail in \cite{Deift,IntSysII,Perel}, for example.
In contrast, CMV matrices are elements of the  unitary  group and hence the natural backdrop for CMV
is that of  Poisson-Lie groups or, more specifically, the group  $Gl(n,\IC)$ equipped with the  Sklyanin bracket
(see \cite[\S 2.12]{IntSysII}).
However, we choose to give a presentation in which the associative algebra of $n\times n$
matrices takes center stage; an analogous construction for KdV using
the algebra of pseudo-differential operators was given by Gelfand
and Dikij \cite{GD}. This approach is described in Section~2.12.6
of \cite{IntSysII}.  (Note that here we are referring to the second
symplectic structure associated with KdV, which was originally
proposed by Adler \cite[\S 4]{Adler}.)

Let $\g$ denote the (associative) algebra of $n\times n$ complex
matrices.  The algebra structure gives rise to a natural Lie algebra
structure:
$$
  [B,C] = BC-CB.
$$
As a vector space, $\g=\LL\oplus\A$, where
$$
\A = \{A : A = - A^\dagger \},
$$
is the space of skew-Hermitian matrices, which is the Lie algebra of the group $\U(n)$ of $n\times n$ unitary
matrices, and
$$
\LL = \{A\in\g : L_{i,j} = 0 \text{ for $i>j$ and } L_{i,i}\in\IR
\}
$$
is the space of upper triangular matrices with real diagonal entries (the Lie algebra of the group $\LLL(n)$ of $n\times n$ lower
triangular matrices with positive diagonal entries). We will write
$\pi_{\A}$ and $\pi_\LL$ for the natural projections onto these summands.
This vector-space splitting of $\g$ gives rise to a second
Lie algebra structure. First we define $R:\g\to\g$ by either
\begin{equation}\label{E:RDef}
\begin{aligned}
R(X) &= \pi_\LL(X) - \pi_\A(X),  &  \qquad&\text{for all $X\in\g$, or} \\
R(L+A)&= L-A    &  \qquad&\text{for all $A\in\A$ and $L\in\LL$.}
\end{aligned}
\end{equation}
The second Lie bracket can then be written as either
\begin{equation}\label{E:DefnRBracket}
\begin{aligned}{}
[X,Y]_{\RR} = \tfrac12 [R(X),Y] + \tfrac12 [X,R(Y)]     \quad&\text{$\forall$ $X,Y\in\g$, or} \\
[L+A,L'+A']_{\RR} = [L,L'] - [A,A']    \quad&\text{$\forall$
$L,L'\in\LL$, and $A,A'\in\A$.}
\end{aligned}
\end{equation}
The second definition also makes it transparent that the bracket $[\
,\ ]_{R}$ obeys the Jacobi identity. But the main property that we
are interested in is that $R$ obeys the modified classical
Yang-Baxter equation:
\begin{equation*}
[R(X),R(Y)] - R\bigl([R(X),Y] + [X,R(Y)]\bigr) = -[X,Y]
\end{equation*}
This allows us to define Poisson brackets as follows:
We can identify the dual space  $\g^*$ with $\g$
using  the pairing
\begin{equation}\label{pairing}
\langle X,Y\rangle = \Im \Tr (XY).
\end{equation}
The form $ \langle\ ,\ \rangle$ is is non-degenerate
symmetric and {\it invariant}:
\begin{equation}
\langle X, [Z,Y] \rangle = \Im \Tr (XZY-XYZ) = \langle
[X,Z],Y\rangle
\end{equation}
or, equivalently,
\begin{equation}
\langle BXB^{-1}, BYB^{-1} \rangle = \langle X, Y
\rangle,\qquad\text{for any $B\in GL(n,\IC)$.}
\end{equation}
Given a smooth function $\varphi:\g\to\IR$ and $B\in\g$, define $\nabla\varphi:\g\to\g$
by
\begin{equation}
\frac{d}{dt}\bigg|_{t=0} \varphi(B+tC) =
\langle\nabla\varphi\big|_B,C\rangle.
\end{equation}
Equivalently, if we write $b_{k,l}=u_{k,l} + i v_{k,l}$ for the
matrix entries of $B$, then
\begin{equation}
 [\nabla\varphi]_{k,l} = \frac{\partial\varphi}{\partial v_{l,k}} + i \frac{\partial\varphi}{\partial u_{l,k} }\ .
\end{equation}

We can now define the desired Poisson brackets on $\g$.
Let $R:\g\to\g$ and $\langle\cdot\,,\cdot\rangle$ be as above.
Given $\varphi_1,\varphi_2:\g\to\IR$, let $\nabla_j=\nabla\varphi_j$ for $j=1,2$.  Then both
\begin{equation}\label{E:KKD}
\{\varphi_1,\varphi_2\}_{_{LP}}\bigr|_X=\tfrac{1}{2} \langle [\na_1,\na_2]_{\RR} ,
X \rangle
\end{equation}
and
\begin{equation}\label{E:GDD}
\{\varphi_1,\varphi_2\}_{_{GD}}\bigr|_X = \tfrac{1}{2} \langle R(\nabla_1X), \nabla_2X \rangle -
\tfrac{1}{2}\langle R(X\nabla_1), X\nabla_2 \rangle
\end{equation}
define Poisson structures on $\g$, the first one known as the Lie-Poisson (LP) bracket, and the second known
as the Gelfand-Dikij (GD) bracket. Then
under the embedding $J\mapsto iJ$, the manifolds of Jacobi matrices
with fixed trace are symplectic leaves in $(\g,\{\cdot,\cdot\}_{_{LP}})$; this is just the usual
construction in $Gl(n,\IR)$ in disguise. Similarly,
the manifold of CMV matrices with fixed determinant forms a
symplectic leaf in the Poisson manifold $(\g,\{\cdot,\cdot\}_{_{GD}})$. Furthermore, the restriction of
the GD Poisson structure to the manifold of CMV matrices coincides with the AL Poisson structure~\eqref{E:PbDefn} (see \cite{KN}).

\section{Poisson brackets for Carath\'eodory and Schur functions}\label{S:3}

Let $\nabla_1$ and $\nabla_2$ be the gradients at a point $X$ of two functions $\varphi_1$ and $\varphi_2$,
as defined in Subsection~\ref{Ss:2.2}.
Then a simple calculation shows that the two brackets defined above can be written in the following slightly modified form:
\begin{equation}\label{E:7}
\{\varphi_1,\varphi_2\}_{_{LP}}
=\tfrac14 \langle [X,\nabla_1] ,R(2\nabla_2)\rangle
-\tfrac14 \langle [X,\nabla_2], R(2\nabla_1)\rangle
\end{equation}
and
\begin{equation}\label{E:Rcomm}
\{\varphi_1,\varphi_2\}_{_{GD}}
=\tfrac14 \langle [X,\nabla_1] ,R(X\nabla_2+\nabla_2 X)\rangle
-\tfrac14 \langle [X,\nabla_2], R(X\nabla_1+\nabla_1 X)\rangle.
\end{equation}
Before going any further, we give the following definition. Consider $k\geq 0$ and two smooth
functions $\varphi_1$ and $\varphi_2$, with gradients $\na_1$ and $\na_2$ respectively. Then we set
\begin{equation}\label{E:DefnkBrac}
4\{\varphi_1,\varphi_2\}^{(k)}
=\Big\langle X,\bigl[\na_1, R(X^k\na_2+\na_2X^k)\bigr]+\bigl[R(X^k\na_1+\na_1X^k),\na_2\bigr]\Big\rangle.
\end{equation}
One must note that this expression is a Poisson bracket only for $k=0$, in which case it is exactly the
Lie-Poisson bracket, and for $k=1$. In this latter case a simple calculation using the properties of $R$ and \eqref{E:Rcomm} shows
that $\{\cdot,\cdot\}^{(1)}$ is actually the Gelfand-Dikij bracket defined in \eqref{E:GDD}. But even though
\eqref{E:DefnkBrac} is not in general a Poisson bracket, in what follows we will work with the general expression
as it makes it easier to emphasize the relevant steps of our calculations.

We now focus on functions given by
$$\varphi(X)=\Im\Tr(P\Phi(X)P)=\Im(\Phi(X)_{11}),$$ where $\Phi$ is a smooth
function on $M_n(\IC)$, and $P$ denotes the orthogonal projection on the vector
$e_1=(1,0,\dots,0)^T$,
$$
P=e_1^T e_1.
$$
Note that any $\varphi$ defined this way is
invariant under the conjugation by invertible matrices with off-diagonal
entries in the first column and row all equal to zero:
$$
\varphi(C X C^{-1})= \varphi(X)$$
for
$$
C = \left ( \begin{array} {cc} c_1 & 0\\ 0 & C_2 \end{array} \right )\ ,
$$
where $c_1 \ne 0$ and $C_2$ is an invertible $(n-1)\times (n-1)$ matrix.

\begin{lemma}\label{L:pairing}
Let $\varphi(X)=\Im (\Phi(X)_{11})$ as above. Then for any matrix $B\in M_n(\IC)$ we have
\begin{equation}\label{E:pairing}
\langle\nabla \varphi\bigr|_X, [X,B]\rangle=
\langle\nabla \varphi\bigr|_X, [X,BP+PB]\rangle=\langle[\nabla \varphi\bigr|_X,X],BP+PB\rangle\,
\end{equation}
where, as above, $P=e_1^T e_1$ is the orthogonal projection on the vector
$e_1=(1,0,\dots,0)^T$.
\end{lemma}

\begin{proof}
Let $C = \exp ( t A) $, where $A$ is any matrix of the same form as $C$.
Then
$$
0= \left.\frac{d} {dt} \right|_{t=0} \varphi(C X C^{-1}) =
\langle \nabla \varphi\bigr|_X, [A, X] \rangle\ ,
$$
which implies that, for any $B$ an expression $\langle \nabla \varphi\bigr|_X, [X, B] \rangle$ does not
depend on the $(n-1)\times(n-1)$ submatrix of $B$ obtained  by deleting the first
row and column.
\end{proof}

Using the properties of the $R$-matrix and
Lemma~\ref{L:pairing} we see that:
\begin{equation*}\label{E:Rbrackk}
\begin{aligned}
4\{\varphi_1,\varphi_2\}^{(k)}
&=\Big\langle \bigl[X,\na_1\bigr], R(X^k\na_2+\na_2X^k)\Big\rangle -\Big\langle \bigl[X,\na_2\bigr], R(X^k\na_1+\na_1X^k)\Big\rangle\\
&=\Big\langle \na_1,\bigl[R(X^k\na_2+\na_2X^k),X\bigr]\Big\rangle -\Big\langle \na_2,\bigl[R(X^k\na_1+\na_1X^k),X\bigr]\Big\rangle\\
&=\Big\langle \na_1,\bigl[PR(X^k\na_2+\na_2X^k)+R(X^k\na_2+\na_2X^k)P,X\bigr]\Big\rangle\\
&\quad-\Big\langle \na_2,\bigl[PR(X^k\na_1+\na_1X^k)+R(X^k\na_1+\na_1X^k)P,X\bigr]\Big\rangle\\
&=\Big\langle \bigl[X,\na_1\bigr],PR(X^k\na_2+\na_2X^k)+R(X^k\na_2+\na_2X^k)P\Big\rangle\\
&\quad-\Big\langle \bigl[X,\na_2\bigr],PR(X^k\na_1+\na_1X^k)+R(X^k\na_1+\na_1X^k)P\Big\rangle\\
\end{aligned}
\end{equation*}
While this expression appears to be more complicated than the one we started from, it will
shortly be shown that it is exactly what we need in order to continue our calculation.

Using Lemma \ref{L:pairing}, we can compute the Poisson bracket for functions $\varphi$ as above. Note that
this calculation is general and extremely robust, and it applies to a large range of expressions
involving $R$-matrices. The first such calculation that we are aware of was done for the finite
Toda lattice by Faybusovich and Gekhtman~\cite{FG2}. Proposition~\ref{P:bracket}
gives a short proof of the next step in the calculation, while Theorem~\ref{T:CSbracket} takes it
to its conclusion.
\begin{prop}\label{P:bracket}
Let $\varphi_j(X)=\Im(\Phi_j(X)_{11})=\Im\Tr (\Phi_j(X)P)$ for $j=1,2$, where $\Phi_j$ are smooth functions,
and set $\na_j=\na \varphi_j(X)$. Then
the values of the linear and quadratic brackets of the $\varphi_j$'s are given by:
\begin{equation}\label{E:bracketLP}
2\{\varphi_1,\varphi_2\}_{_{LP}}\bigr|_X
=\Im\Bigl(2(X\na_1\na_2^*)_{11}-2(X\na_2\na_1^*)_{11}+\bigl[X,[\na_1,\na_2]\bigr]_{11}\Bigr)
\end{equation}
and
\begin{equation}\label{E:bracket0}
\begin{aligned}
2\{\varphi_1,\varphi_2\}_{_{GD}}\bigr|_X
=\Im\Bigl(&2(X\na_1\na_2^*X^*)_{11}-2(\na_1XX^*\na_2^*)_{11}\\
&+[X\na_1,X\na_2]_{11}-[\na_1X,\na_2X]_{11}\Bigr).
\end{aligned}
\end{equation}
\end{prop}

\begin{proof}
First note that for any matrix $A$ we have
$$
PR(A)=2\Re(A_{11})P-PA\quad\text{and}\quad
R(A)P=AP+2A^*P-2\Re(A_{11})P.
$$
Therefore
$$
PR(A)+R(A)P=[A,P]+2A^*P
$$
and so for any $k\geq0$ the expression $4\{\varphi_1,\varphi_2\}^{(k)}$ equals
\begin{equation}\label{E:brackk2}
\begin{aligned}
&\Big\langle \big[X,\na_1\big],2\big(X^k\na_2+\na_2X^k\big)^*P\Big\rangle
-\Big\langle \big[X,\na_2\big],2\big(X^k\na_1+\na_1X^k\big)^*P\Big\rangle\\
&+\Big\langle \big[X,\na_1\big],\big[X^k\na_2+\na_2X^k,P\big]\Big\rangle
-\Big\langle \big[X,\na_2\big],\big[X^k\na_1+\na_1X^k,P\big]\Big\rangle
\end{aligned}
\end{equation}
Taking \eqref{E:brackk2} for $k=0$ and 1 yields the result.
\end{proof}

Note that the $R$-matrix does not appear any more in \eqref{E:bracketLP} or \eqref{E:bracket0}.
This allows us to prove our first main result: compute the respective brackets for the
Weyl and Carath\'eodory functions. More precisely, let $X\in M_n(\IC)$ be a matrix, and $z, \la \in \IC\setminus \spec(X)$
complex numbers. For $k\geq0$, consider
\begin{equation}
 m^{(k)}(\la)=\bigl(X^k(\la-X)^{-1} \bigr)_{11}=\Tr\bigl(X^k(\la-X)^{-1}P\bigr)
\end{equation}
and
\begin{equation}
 F^{(k)}(z)=\bigl(X^k(X+z)(X-z)^{-1}\bigr)_{11}=\Tr\bigl(X^k(X+z)(X-z)^{-1}P\bigr).
\end{equation}
We think of these functions as being defined on the space of matrices, and depending on a complex parameter,
$\la$ or $z$ respectively. Note that, for $k=0$, we recover the Weyl and Carath\'eodory functions, respectively.
We then have the following result:
\begin{theorem}\label{T:CSbracket}
Let $X\in M_n(\IC)$ be a fixed matrix, and $z,w,\la,\xi \in \IC\setminus \spec(X)$ be complex parameters,
with $z\neq w$ and $\la\neq \xi$. Then, for $k\geq0$,
\begin{equation}\label{E:10}
\{m(\la),m(\xi)\}^{(k)}=\bigl(m^{(k)}(\la)-m^{(k)}(\xi)\bigr)
\left[m(\la)m(\xi)-\frac{m(\la)-m(\xi)}{\la-\xi}\right]
\end{equation}
and
\begin{equation}\label{E:11}
\{F(z),F(w)\}^{(k+1)}=i\bigl(F^{(k)}(z)-F^{(k)}(w)\bigr)
\left[F(z)F(w)-1-\frac{z+w}{z-w}\bigl(F(z)-F(w)\bigr)\right].
\end{equation}
\end{theorem}

\begin{coro}\label{C:3.1}
Let $\C$ be a finite CMV matrix, and $F$ and $f$ the Carath\'eodory and Schur functions
associated to the spectral measure of $\C$ and $e_1=[1,0,\ldots,0]^T$. Then, for two distinct points
$z,w\in \IC\setminus \spec(\C)$, the GD brackets
of these functions at $\C$ are given by:
\begin{equation}\label{E:Cbrac}
\begin{aligned}
\{F(z),F(w)\}_{_{GD}}
&=i\bigl(F(z)-F(w)\bigr)\bigl(F(z)F(w)-1\bigr)\\
&\quad-i\frac{z+w}{z-w}\bigl(F(z)-F(w)\bigr)^2
\end{aligned}
\end{equation}
and
\begin{equation}\label{E:Sbrac}
\{f(z),f(w)\}_{_{GD}}=-2i\frac{f(z)-f(w)}{z-w}\bigl(zf(z)-wf(w)\bigr).
\end{equation}
\end{coro}

\begin{proof}
The first relation is just a special case of \eqref{E:11} for $k=0$ and $X=\C$.
The bracket \eqref{E:Sbrac} follows from \eqref{E:CaraSchur} and the observation that
\begin{align*}
\{f(z),f(w)\}_{_{GD}}
&=\frac{1}{zw}\, \frac{d}{dF(z)}\left(\frac{1-F(z)}{1+F(z)}\right) \frac{d}{dF(w)}\left(\frac{1-F(w)}{1+F(w)}\right)
\{F(z),F(w)\}_{_{GD}}\\
&=\frac{4}{zw}\, \frac{1}{(1+F(z))^2(1+F(w))^2}\{F(z),F(w)\}_{_{GD}}.
\end{align*}
Now use \eqref{E:Cbrac} and the expression \eqref{E:CaraSchur} of $F$ in terms of $f$.
\end{proof}

\begin{proof}[Proof of Theorem~\ref{T:CSbracket}]
The proofs of both relations follow the exact same ideas, but since in this paper we focus on the Ablowitz-Ladik system,
and hence the Gelfand-Dikij bracket, we will only give the proof of \eqref{E:11}. We approach this by first computing
the $k$-brackets of $\Im F$ and $\Re F=\Im (iF)$
at two different points $z$ and $w$.

Let $R$ be the R-matrix defined in Section~\ref{S:2}. We start by working with the general
expression for the $k$-bracket and for any two functions
$\varphi_1$ and $\varphi_2$ as above. We know that
\begin{equation*}
\begin{aligned}
4\{\varphi_1,\varphi_2\}^{(k)}(X)
&=\Big\langle \bigl[X,\na_1\bigr],PR(X^k\na_2+\na_2X^k)+R(X^k\na_2+\na_2X^k)P\Big\rangle\\
&\quad-\Big\langle \bigl[X,\na_2\bigr],PR(X^k\na_1+\na_1X^k)+R(X^k\na_1+\na_1X^k)P\Big\rangle\\
\end{aligned}
\end{equation*}
Just as in the proof of Proposition~\ref{P:bracket}, the observation that allows us to continue is that, for any matrix $A$:
$$
PR(A)=2\Re(A_{11})P-PA\quad\text{and}\quad
R(A)P=AP+2A^*P-2\Re(A_{11})P.
$$
Therefore
$$
PR(A)+R(A)P=[A,P]+2A^*P
$$
and so
\begin{equation}
\begin{aligned}
4\{\varphi_1,\varphi_2\}^{(k)}(X)
&=\Big\langle \big[X,\na_1\big],2\big(X^k\na_2+\na_2X^k\big)^*P\Big\rangle\\
&\quad -\Big\langle \big[X,\na_2\big],2\big(X^k\na_1+\na_1X^k\big)^*P\Big\rangle\\
&\quad+\Big\langle \big[X,\na_1\big],\big[X^k\na_2+\na_2X^k,P\big]\Big\rangle\\
&\quad -\Big\langle \big[X,\na_2\big],\big[X^k\na_1+\na_1X^k,P\big]\Big\rangle
\end{aligned}
\end{equation}

We wish to apply this formula to the real and imaginary parts of the Carath\'eodory function $F$
$$
F(z)=\left(\frac{X+z}{X-z}\right)_{11}.
$$
Indeed, let $u(z)=\Re F(z)=\Im iF(z)$ and $v(z)=\Im F(z)$.
Note that, for a fixed parameter $z$, these functions are well-defined in a neighborhood of $X$, and they are of
the type we have considered above. If we denote $\na_z=\na v(z)$, then we get
$$
\na u(z)=i\na_z=-2iz(X-z)^{-1}P(X-z)^{-1}
$$
and
$$
\na v(z)=\na_z=-2z(X-z)^{-1}P(X-z)^{-1}.
$$
Plugging these into the formula for the $k^\text{th}$ $R$-bracket, we get
\begin{eqnarray*}
\{u(z),u(w)\}^{(k)} & = & \Im\bigl(T_1(k)-T_2(k)-T_3(k)+T_4(k)\bigr)\\
\{v(z),v(w)\}^{(k)} & = & \Im\bigl(T_1(k)-T_2(k)+T_3(k)-T_4(k)\bigr)\\
\{u(z),v(w)\}^{(k)} & = & \Im\bigl(iT_1(k)+iT_2(k)+iT_3(k)-iT_4(k)\bigr)\\
  & = & \Re\bigl(T_1(k)+T_2(k)+T_3(k)-T_4(k)\bigr)\\
\{v(z),u(w)\}^{(k)} & = & \Im\bigl(-iT_1(k)-iT_2(k)+iT_3(k)-iT_4(k)\bigr)\\
  & = & \Re\bigl(-T_1(k)-T_2(k)+T_3(k)-T_4(k)\bigr),
\end{eqnarray*}
where
\begin{eqnarray*}
T_1(k) & = & \frac14\Tr \Bigl(2\big[X,\na_z\big]\big(X^k\na_w+\na_wX^k\big)^*P\Bigr)\\
T_2(k) & = & \frac14\Tr \Bigl(2\big[X,\na_w\big]\big(X^k\na_z+\na_zX^k\big)^*P\Bigr)\\
T_3(k) & = & \frac14\Tr \Bigl(\big[X,\na_z\big]\cdot\big[X^k\na_w+\na_wX^k,P\big]\Bigr)\\
T_4(k) & = & \frac14\Tr \Bigl(\big[X,\na_w\big]\cdot\big[X^k\na_z+\na_zX^k,P\big]\Bigr).
\end{eqnarray*}
Therefore
\begin{equation*}
\begin{aligned}
\{F(z),F(w)\}^{(k)}
&=\{u(z),u(w)\}^{(k)}-\{v(z),v(w)\}^{(k)}\\
&\quad+i\big(\{u(z),v(w)\}^{(k)}+\{v(z),u(w)\}^{(k)}\big)\\
&=-2\Im\big(T_3(k)-T_4(k)\big)+i\cdot2\Re\big(T_3(k)-T_4(k)\big)\\
&=2i\big(T_3(k)-T_4(k)\big)
\end{aligned}
\end{equation*}
Let us note in passing that in the $k=1$ (Gelfand-Dikij) case, we have that $T_1=T_3$ and $T_2=T_4$.

In order to compute $T_3-T_4$, we also need the following simple observation: For any three matrices $A,B$ and $C$,
\begin{equation}\label{E:Formula1}
\Tr\bigl(APBPCP\bigr)=\Tr\bigl(PAPBPCP\bigr)=A_{11}B_{11}C_{11}.
\end{equation}
After simplifying and grouping terms together, we get that
$$
4T_3(k)-4T_4(k)={\rm I}(k)+{\rm II}(k)+{\rm III}(k),
$$
where
\begin{equation*}
\begin{aligned}
{\rm I}(k)=8zw\bigl((X-z)^{-1}X^{k}(X-w)^{-1}\bigr)_{11}\cdot
&\Bigl[\bigl(X(X-z)^{-1}\bigr)_{11}\bigl((X-w)^{-1}\bigr)_{11}\\
&\quad-\bigl(X(X-w)^{-1}\bigr)_{11}\bigl((X-z)^{-1}\bigr)_{11}\Bigr],\\
\end{aligned}
\end{equation*}
\begin{equation*}
\begin{aligned}
{\rm II}(k)=8zw\bigl((X-z)^{-1}X(X-w)^{-1}\bigr)_{11}\cdot
&\Bigl[\bigl(X^{k}(X-z)^{-1}\bigr)_{11}\bigl((X-w)^{-1}\bigr)_{11}\\
&\quad-\bigl(X^{k}(X-w)^{-1}\bigr)_{11}\bigl((X-z)^{-1}\bigr)_{11}\Bigr],\\
\end{aligned}
\end{equation*}
and
\begin{equation*}
\begin{aligned}
{\rm III}(k)=8zw\bigl((X-z)^{-1}(X-w)^{-1}\bigr)_{11}\cdot
&\Bigl[\bigl(X(X-z)^{-1}\bigr)_{11}\bigl(X^{k}(X-w)^{-1}\bigr)_{11}\\
&\quad-\bigl(X(X-w)^{-1}\bigr)_{11}\bigl(X^{k}(X-z)^{-1}\bigr)_{11}\Bigr].\\
\end{aligned}
\end{equation*}
Recall that, for $j\geq 0$, we have defined the function
\begin{equation}
F^{(j)}(z)=\left(X^j\frac{X+z}{X-z}\right)_{11},\qquad j\in\ZZ.
\end{equation}
Then we have that, for $j\in\ZZ$,
\begin{equation}\label{E:1}
\bigl(X^{j+1}(X-z)^{-1}\bigr)_{11}
=\frac12\bigl(F^{(j)}(z)+F^{(j)}(0)\bigr)
=\frac{1}{2z}\bigl(F^{(j+1)}(z)-F^{(j+1)}(0)\bigr),
\end{equation}
\begin{equation}\label{E:2}
\bigl((X-z)^{-1}X^{j+1}(X-w)^{-1}\bigr)_{11}
=\frac{1}{2(z-w)}\bigl(F^{(j)}(z)-F^{(j)}(w)\bigr),
\end{equation}
and
\begin{equation}\label{E:3}
\bigl((X-z)^{-1}(X-w)^{-1}\bigr)_{11}
=\frac{1}{2(z-w)}\Bigl(\frac1z\bigl(F(z)-1\bigr)-\frac1w\bigl(F(w)-1\bigr)\Bigr).
\end{equation}

We will use the appropriate formula of \eqref{E:1}--\eqref{E:3} in order to express our
quantities only in terms of $F\equiv F^{(0)}$ and $F^{(k)}$. Thus we get
\begin{equation}
\begin{aligned}
(z-w)\cdot{\rm I}(k+1)=\bigl(F^{(k)}(z)-F^{(k)}(w)\bigr)
&\big[z(F(z)+1)(F(w)-1)\\
&\quad-w(F(w)+1)(F(z)-1) \big],
\end{aligned}
\end{equation}
\begin{equation}
\begin{aligned}
(z-w)\cdot{\rm II}(k+1)=\bigl(F(z)-F(w)\bigr)
&\big[z(F^{(k)}(z)+F^{(k)}(0))(F(w)-1)\\
&\quad-w(F^{(k)}(w)+F^{(k)}(0))(F(z)-1) \big],
\end{aligned}
\end{equation}
and
\begin{equation}
\begin{aligned}
(z-w)\cdot{\rm III}(k+1)&=\bigl(w(F(z)-1)-z(F(w)-1)\bigr)\times\\
&\quad\big[(F(z)+1)(F^{(k)}(w)+F^{(k)}(0))
-(F(w)+1)(F^{(k)}(z)+F^{(k)}(0)) \big].
\end{aligned}
\end{equation}

A straightforward calculation shows that ${\rm II}+{\rm III}={\rm I}$, and hence we find that
\begin{equation}
\{F(z),F(w)\}^{(k+1)}=i\bigl(F^{(k)}(z)-F^{(k)}(w)\bigr)
\Bigl[F(z)F(w)-1-\frac{z+w}{z-w}\bigl(F(z)-F(w)\bigr)\Bigr],
\end{equation}
as claimed.
\end{proof}

We wish to use \eqref{E:Cbrac} to find the bracket of the eigenvalues and masses of the spectral measure,
$\mu=\sum_{j=1}^n\delta_{z_j}\mu_j$,
$z_j=e^{i\theta_j}$ and $\sum\mu_j=1$,
of a CMV matrix $\C$. Note that, since for the next couple of sections we only work with the GD bracket,
we will not specify it in order to simplify notation. We hope that this will not cause any confusion.

Fix a CMV matrix $\C$, with $d\mu=\sum_{j=1}^n\delta_{z_j}\mu_j$,
$z_j=e^{i\theta_j}$ and $\sum\mu_j=1$, the associated spectral measure.
Let us expand the bracket $\{F(z),F(w)\}\big|_\C$ in terms of the $z_j$s and $\mu_j$s:
\begin{align*}
\{F(z),F(w)\}
&=\sum_{j,k=1}^n \left\{\frac{z+z_j}{z-z_j}\mu_j,\frac{w+z_k}{w-z_k}\mu_k\right\}\\
&=\sum_{j,k=1}^n \frac{z+z_j}{z-z_j}\frac{w+z_k}{w-z_k}\{\mu_j,\mu_k\}\\
&\quad+\sum_{j,k=1}^n \mu_j\frac{w+z_k}{w-z_k}\frac{d}{dz_j}\left(\frac{z+z_j}{z-z_j}\right)\{z_j,\mu_k\}\\
&\quad+\sum_{j,k=1}^n \mu_j\mu_k\frac{d}{dz_j}\left(\frac{z+z_j}{z-z_j}\right)
\frac{d}{dz_k}\left(\frac{w+z_k}{w-z_k}\right)\{z_j,z_k\}.\\
\end{align*}
Since
$$
\frac{d}{d\ze}\left(\frac{z+\ze}{z-\ze}\right)=\frac{2z}{(z-\ze)^2}\,,
$$
we obtain
\begin{align*}
\{F(z),F(w)\}
&=\sum_{j,k=1}^n \frac{(z+z_j)(w+z_k)}{(z-z_j)(w-z_k)}\cdot\{\mu_j,\mu_k\}\\
&\quad+ \sum_{j,k=1}^n \frac{2z(w+z_k)\mu_j}{(z-z_j)^2(w-z_k)}\cdot\{z_j,\mu_k\}\\
&\quad+ \sum_{j,k=1}^n \frac{4zw\mu_j\mu_k}{(z-z_j)^2(w-z_k)^2}\cdot\{z_j,z_k\}.
\end{align*}

Let $1\leq s,t\leq n$. If $s\neq t$, then we choose $\Gamma_s$ and $\Gamma_t$ to be small,
positively oriented contours around $z_s$ and $z_t$, respectively;
we require that they do not intersect, nor surround more that one eigenvalue.
In the case $s=t$, we choose two small, positively oriented contours $\Gamma_s$ and $\Gamma_s^\prime$
around $z_s$ so that the contour $\Gamma_s$ is completely contained in the interior of $\Gamma_s^\prime$.
By the residue formula the previous expansion
implies that
\begin{equation}\label{E:bracmm}
\frac{1}{(2\pi i)^2}\int_{\Gamma_t}\int_{\Gamma_s}\{F(z),F(w)\}\,dzdw
=4z_sz_t\{\mu_s,\mu_t\},
\end{equation}
\begin{equation}\label{E:braczm}
\frac{1}{(2\pi i)^2}\int_{\Gamma_t}\int_{\Gamma_s}(z-z_s)\{F(z),F(w)\}\,dzdw
=4z_sz_t\mu_s\{z_s,\mu_t\},
\end{equation}
and
\begin{equation}\label{E:braczz}
\frac{1}{(2\pi i)^2}\int_{\Gamma_t}\int_{\Gamma_s}(z-z_s)(w-z_t)\{F(z),F(w)\}\,dzdw
=4z_sz_t\mu_s\mu_t\{z_s,z_t\}.
\end{equation}
Note that, if $s=t$, then we set $\Gamma_t=\Gamma_s^\prime$; in other words, we first integrate over the smaller of the
two contours around $z_s$.

\begin{theorem}\label{T:bracket}
With the definitions from the previous sections we have that, in the GD Poisson structure,
\begin{equation}\label{E:braczz2}
\{z_s,z_t\}=0,
\end{equation}
\begin{equation}\label{E:braczm2}
\{z_s,\mu_t\}=2iz_s\mu_t(\delta_{st}-\mu_s),
\end{equation}
and
\begin{equation}\label{E:bracmm2}
\{\mu_s,\mu_t\}=2i\mu_s\mu_t \biggl[\sum_{k\neq s}\frac{z_k+z_s}{z_k-z_s}\mu_k
+ \sum_{k\neq t}\frac{z_t+z_k}{z_t-z_k}\mu_k  + \frac{z_s+z_t}{z_s-z_t}\biggr].
\end{equation}
\end{theorem}

\begin{proof}
We prove the result by finding the residues generated by the right-hand side of
\begin{equation}\label{E:bracket2.1}
\begin{aligned}
\{F(z),F(w)\}=
&i\bigl(F(z)-F(w)\bigr)\bigl(F(z)F(w)-1\bigr)\\
&\,-i\frac{z+w}{z-w}\bigl(F(z)-F(w)\bigr)^2
\end{aligned}
\end{equation}
in the integrals from \eqref{E:bracmm},\eqref{E:braczm}, and \eqref{E:braczz}.

We begin with \eqref{E:braczz}. Note that only the quadratic poles in
both $z-z_j$ and $w-z_k$ play any role. But for $s\neq t$ there is no term on the right-hand side
of \eqref{E:bracket2.1} which contains the denominator $(z-z_s)^2(w-z_t)^2$, and hence the
double integral over $\Gamma_t$ and $\Gamma_s$ is identically 0. This proves \eqref{E:braczz2}.

Now we turn to \eqref{E:braczm}. Assume first that $s\neq t$. By the same reasoning than above, the
only terms in \eqref{E:bracket2.1} which contribute to the integral are the ones containing the denominator
$(z-z_s)^2(w-z_t)$. In this case, that translates into
\begin{equation*}
\begin{aligned}
4z_sz_t\mu_s\{z_s,\mu_t\}
&=\frac{1}{(2\pi i)^2}\int_{\Gamma_t}\int_{\Gamma_s}(z-z_s)iF(z)^2F(w)\,dzdw\\
&=\frac{1}{(2\pi i)^2}\int_{\Gamma_t}\int_{\Gamma_s}i(z-z_s)\frac{(z+z_s)^2}{(z-z_s)^2}\mu_s^2
\frac{w+z_t}{z_t-w}\mu_t\,dzdw\\
&=(-i)4z_s^2\mu_s^22z_t\mu_t\, ,
\end{aligned}
\end{equation*}
or, equivalently,
\begin{equation}\label{E:braczm1}
\{z_s,\mu_t\}=-2iz_s\mu_s\mu_t
\end{equation}
for $s\neq t$. The case $s=t$ is slightly more complicated because the factor $(z+w)/(z-w)$
in the second term of the right-hand side of
\eqref{E:bracket2.1} plays a role. Indeed,
$$
\frac{1}{2\pi i}\int_{\Gamma_s} (z-z_s)\{F(z),F(w)\}\,dz
=i4z_s^2\mu_s^2F(w)-i\frac{z_s+w}{z_s-w}4z_s^2\mu_s^2.
$$
Recall that $\Gamma_s$ is contained in the interior of $\Gamma_s^\prime$, and hence the function
$z\to\frac{z+w}{z-w}$ is analytic on an open neighborhood of the interior of $\Gamma_s$ as long
as $w$ is on $\Gamma_s^\prime$. Now integrate on $\Gamma_s^\prime$:
\begin{equation*}
4z^2_s\mu_s\{z_s,\mu_s\}=-8iz_s^2\mu_s^2z_s\mu_s-i(-2z_s)4z_s^2\mu_s^2=4z_s^2\mu_s(-2iz_s\mu_s^2+2iz_s\mu_s).
\end{equation*}
Combining this with \eqref{E:braczm1} gives \eqref{E:braczm2}.

Finally, we turn to \eqref{E:bracmm}. Here the formulae are more involved and all the terms play a role.
\begin{align*}
\frac{1}{2\pi i}\int_{\Gamma_s}\{F(z),F(w)\}\,dz
&=-2iz_s\mu_s \biggl(F(w)\sum_{k\neq s}\frac{z_s+z_k}{z_s-z_k}\mu_k-1\biggr)\\
&\quad-2iz_s\mu_sF(w) \biggl(\sum_{k\neq s}\frac{z_s+z_k}{z_s-z_k}\mu_k -F(w)\biggr)\\
&\quad -4iz_s\mu_s\frac{z_s+w}{z_s-w}\biggl(\sum_{k\neq s}\frac{z_s+z_k}{z_s-z_k}\mu_k-F(w)\biggr).
\end{align*}
Integrating this on $\Gamma_t$ for $t\neq s$ yields
\begin{align*}
4z_sz_t\{\mu_s,\mu_t\}
&=-8iz_sz_t\mu_z\mu_t\sum_{k\neq s}\frac{z_s+z_k}{z_s-z_k}\mu_k\\
&\quad+8iz_sz_t\mu_s\mu_t\sum_{k\neq t}\frac{z_t+z_k}{z_t-z_k}\mu_k\\
&\quad+8iz_sz_t\mu_s\mu_t\frac{z_s+z_t}{z_s-z_t},
\end{align*}
or, after simplifications,
\begin{align*}
\{\mu_s,\mu_t\}
=2\mu_s\mu_t \biggl[-\sum_{k\neq s}i\frac{z_s+z_k}{z_s-z_k}\mu_k + \sum_{k\neq t}i\frac{z_t+z_k}{z_t-z_k}\mu_k
+ i\frac{z_s+z_t}{z_s-z_t}\biggr],
\end{align*}
as claimed.
\end{proof}

\begin{remark}
As an immediate consequence of Theorem~\ref{T:bracket} we can recover the analogous formulae
for the Ablowitz-Ladik system obtained by Killip and Nenciu, \cite{KN}. Indeed, let us rewrite the
brackets in terms of $\theta_s$ (where $z_s=e^{i\theta_s}$) and $\log[\mu_t]$. We easily obtain from
\eqref{E:braczz2} and \eqref{E:braczm2} that
$$
\{\theta_s,\theta_t\}=0
$$
and
$$
\{\theta_s,\log[\mu_t]\}=2\delta_{st}-2\mu_s.
$$
The first formula is a direct consequence of the Lax pairs in \cite[Proposition~4.5]{KN}, while the
second one coincides with equation~(72) from the same paper. Consider \eqref{E:bracmm2} and use the
fact that
$$
i\frac{e^{i\varphi_1}+e^{i\varphi_2}}{e^{i\varphi_1}-e^{i\varphi_2}}=\cot\Bigl(\frac{\varphi_1-\varphi_2}{2}\Bigr).
$$
Then we get that
\begin{align*}
\{\log[\mu_s],\log[\mu_t]\}
&=\sum_{k\neq s}\cot\Bigl(\frac{\theta_k-\theta_s}{2}\Bigr)\mu_k
+\sum_{k\neq t} \cot\Bigl(\frac{\theta_t-\theta_k}{2}\Bigr)\mu_k + \cot\Bigl(\frac{\theta_s-\theta_t}{2}\Bigr)\\
&=\sum_{k\neq s,t}
\biggl[\cot\Bigl(\frac{\theta_k-\theta_s}{2}\Bigr)
+\cot\Bigl(\frac{\theta_t-\theta_k}{2}\Bigr)
+ \cot\Bigl(\frac{\theta_s-\theta_t}{2}\Bigr)\biggr]\mu_k.
\end{align*}
In the last identity we use the fact that
$$1-\mu_s-\mu_t=\sum_{k\neq s,t}\mu_k.$$
Note that we recovered Proposition~8.5 from \cite{KN}.

While our proof of \eqref{E:braczz2} and \eqref{E:braczm2} is not necessarily shorter than the one from
\cite{KN}, the case of \eqref{E:bracmm2} is completely different: not only is this derivation much simpler and shorter, it
also only uses the $R$-matrix formulation of the Gelfand-Dikij bracket. The approach of Killip and Nenciu exploits the
asymptotics of the spectral parameters, and the expression of the bracket in terms of Verblunsky coefficients.
\end{remark}

\section{The extended bracket for unnormalized measures}\label{S:ExtendedPB}

We will now extend the bracket from probability measures $\mu$ on the unit circle $S^1$ to general, finite measures $\mut$ on $S^1$.
If $c=\mut(S^1)$ is the total weight, and $F$ the Carath\'eodory function
associated to the normalized measure, note that $c$ and $F$ fully characterize the unnormalized
measure. In other words, in order to define the extended bracket it is sufficient to give its value for these
two quantities:
\begin{equation}\label{E:cbrack}
\{c, F(z)\}=ic(F(z)^2-1),
\end{equation}
and then extend it to all smooth functions using the Leibnitz rule and bi-linearity.\footnote{Note that here we use the same notation
for the Poisson bracket on the space of finite measures as for the one on the space of probability measures. But since the
later is just the restriction of the former, we trust that this will not create any confusion.}
While this describes the bracket uniquely, it is not clear that it obeys the Jacobi identity.
In order to show this, we will rewrite the bracket in a different set of coordinates. Indeed, from
\eqref{E:cbrack} we get, by the same methods in the proof of
Theorem~\ref{T:bracket}, that
$$
\{c,z_s\}=-2icz_s\mu_s
\quad\text{and}\quad
\{c,\mu_s\}=-2ic\mu_s\sum_{j\neq s}\frac{z_s+z_j}{z_s-z_j},
$$
where for a finite measure $\tilde\mu$ we always denote by $\mu$ the
associated probability measure, $\mu=\mut/|\mut|$ and, as in the
previous sections, $\mu=\sum_{j=1}^n \mu_j\delta_{z_j}$,
$z_j=e^{i\theta_j}$. Further set $\mut_j=c\mu_j$ to be the
corresponding weights for the unnormalized measure
$\mut$.\footnote{Recall that any symbol with a tilde refers to the
unnormalized measures.} Note that the space of measures $\mut$ is
$2n$-dimensional and parameterized by
$\theta_1,\ldots,\theta_n,\mut_1,\ldots,\mut_n$. In these variables,
a direct calculation using (\ref{E:braczz2})-(\ref{E:bracmm2}) and
(\ref{E:cbrack})  shows that for all $1\leq s,t\leq n$ the bracket
is given by
\begin{equation}\label{E:ExtBrac1}
\{\theta_s,\theta_t\}=0,\qquad \{\theta_s,\mut_t\}=2\mut_t\delta_{st},
\end{equation}
and, for $s\neq t$,
\begin{equation}\label{E:ExtBrac2}
\{\mut_s,\mut_t\}=2i\mut_s\mut_t\frac{z_s+z_t}{z_s-z_t}=2\mut_s\mut_t\cot\left(\frac{\theta_s-\theta_t}{2}\right).
\end{equation}
So now we can easily prove that
\begin{prop}\label{P:btilde}
The bracket given by \eqref{E:ExtBrac1} and \eqref{E:ExtBrac2} on the space of finite measures supported
at $n$ points on the unit circle obeys the Jacobi identity, and hence it defines a Poisson bracket. Furthermore,
this bracket is nondegenerate.
\end{prop}
\begin{proof}
Checking the Jacobi identity is actually a simple calculation if one uses the following (slight) variations of \eqref{E:ExtBrac1}
and \eqref{E:ExtBrac2} obtained if we replace the $\mut_j$s by $\frac12\log[\mut_j]$:
$$
\{\theta_s,\theta_t\}=0,\qquad \{\theta_s,\tfrac12\log[\mut_t]\}=\delta_{st},
$$
and
$$
\{\tfrac12\log[\mut_s],\tfrac12\log[\mut_t]\}=\tfrac12\cot\left(\tfrac{\theta_s-\theta_t}{2}\right).
$$

The nondegeneracy of the bracket follows immediately from $\{\theta_s,\mut_t\}=2\mut_t\delta_{st}$. Indeed,
let $f$ be a smooth function which is not constant. In that case, $f$ depends nontrivially on at least one
of the variables, say $\theta_1$, and hence $\frac{\partial f}{\partial\theta_1}\neq 0$ on some open set. Therefore
$$
\{f,\mut_1\}=\frac{\partial f}{\partial\theta_1}\{\theta_1,\mut_1\}
=2\mut_1\frac{\partial f}{\partial\theta_1}\neq0
$$
on that open set.
In other words, constants are the only functions which commute with every other function, and hence the Poisson bracket
is nondegenerate.
\end{proof}

On this space, we are able to find the canonical coordinates for the Poisson structure defined above:
\begin{theorem}\label{T:UnnormCC}
Define
\begin{equation}\label{E:Defnq}
q_s=\mut_s\prod_{j\neq s}|z_s-z_j|,\qquad\text{for all}\quad 1\leq s\leq n.
\end{equation}
Then
\begin{equation}\label{E:UnnormPBCC}
\{\theta_s,q_t\}=2q_t\delta_{st}\quad\text{and}\quad\{\theta_s,\theta_t\}=\{q_s,q_t\}=0.
\end{equation}
In other words,
\begin{equation}\label{E:UnnormCC}
\theta_1,\ldots,\theta_n,\frac12\log[q_1],\ldots,\frac12\log[q_n]
\end{equation}
are canonical coordinates on the space of un-normalized measures $\mut$.
\end{theorem}

\begin{proof}
Having guessed the correct quantities which give the canonical coordinates,
the proof is merely a calculation. As all the $z_j$'s Poisson commute, the first bracket
in \eqref{E:UnnormPBCC} follows immediately from \eqref{E:ExtBrac1}.

So we need to compute the bracket of the $q_s$'s. For $s\neq t$ we have
\begin{equation*}
\begin{aligned}
\{q_s,q_t\}
&=\{\mut_s,\mut_t\}\cdot \prod_{j\neq s} |z_j-z_s| \prod_{k\neq t}|z_k-z_t|\\
&\quad +\mut_t\prod_{j\neq s}|z_j-z_s|\cdot\{\mut_s,\prod_{k\neq t}|z_k-z_t|\}\\
&\quad -\mut_s\prod_{k\neq t}|z_k-z_t|\cdot\{\mut_t,\prod_{j\neq s}|z_j-z_s|\}\\
&=\bigl(\{\mut_s,\mut_t\}+T(s,t)-T(t,s)\bigr)\cdot  \prod_{j\neq s} |z_j-z_s| \prod_{k\neq t}|z_k-z_t|,
\end{aligned}
\end{equation*}
where, using \eqref{E:ExtBrac1} and the fact that $s\neq t$, we find that:
\begin{equation*}
\begin{aligned}
T(s,t)
&=\mut_t\sum_{k\neq t}\frac{\{\mut_s,|z_k-z_t|\}}{|z_k-z_t|}\\
&=\frac{\mut_t}{2|z_s-z_t|^2}\cdot\bigl(\{\mut_s,z_s\}(\bar z_s-\bar z_t)+\{\mut_s,\bar z_s\}(z_s-z_t)\bigr)\\
&=i\mut_s\mut_t\frac{z_s\bar z_t-\bar z_s z_t}{|z_s-z_t|^2}.
\end{aligned}
\end{equation*}
By \eqref{E:ExtBrac2}, the antisymmetry of $T$ and using the fact the $|z_j|=1$ for every $1\leq j\leq n$, we find that
\begin{equation*}
\begin{aligned}
\{\mut_s,\mut_t\}+T(s,t)-T(t,s)
&=2i\mut_s\mut_t\left[\frac{z_s+z_t}{z_s-z_t}+\frac{z_s\bar z_t-\bar z_s z_t}{|z_s-z_t|^2}\right]=0,
\end{aligned}
\end{equation*}
and hence
$$
\{q_s,q_t\}=0
$$
for any $1\le s,t\leq n$. This proves the statement of the theorem.
\end{proof}

Once we have the result above, the canonical coordinates for the space of normalized measures, or, equivalently,
of CMV matrices, follow from a simple observation:
\begin{coro}\label{C:NormCC}
With the notations from Theorem~\ref{T:UnnormCC} we get that
\begin{equation}\label{E:NormCC}
\theta_1,\ldots,\theta_{n-1},\tfrac12\log[r_{1,n}],\ldots,\tfrac12\log[r_{n-1,n}]
\end{equation}
are canonical coordinates on the space of CMV matrices with fixed determinant,
where
\begin{equation}\label{E:NormAngl}
r_{j,k}=\frac{\mu_j}{\mu_k}\prod_{l\neq j,k}\left|\frac{z_l-z_j}{z_l-z_k}\right|.
\end{equation}
\end{coro}

\begin{proof}
The observation that justifies our claim completely is
$$
r_{j,k}=\frac{q_j}{q_k}.
$$
Indeed, while each $q_j$ depends on the normalization through $\mut_j=c\mu_j$, their ratios do not:
$$
\frac{\mut_j}{\mut_k}=\frac{c\mu_j}{c\mu_k}=\frac{\mu_j}{\mu_k}.
$$
Having observed this, the claim that \eqref{E:NormCC} are canonical coordinates
follows by a moment's reflection from \eqref{E:UnnormPBCC}.
\end{proof}

\section{Compatible Poisson brackets}\label{S:CompatiblePB}

In this section, we define a family of compatible (in the sense of Magri) Poisson brackets on the space of
finite measures on the unit circle. Furthermore, the restrictions of all of these brackets to the manifold
of probability measures represents a multi-Hamiltonian structure for the Ablowitz-Ladik equation \eqref{ALE2} described
in the Introduction.

Let $h$ be a smooth function on $\IC$ which takes real values on $S^1$. Define $\{\cdot,\cdot\}_h$
by specifying the bracket of the coordinates $\theta_s$ and $q_t$, $1\leq s,t\leq n$:
\begin{equation}\label{E:DefPBh}
\{\theta_s,\theta_t\}_h=\{q_s,q_t\}_h=0\quad \text{and}\quad
\{\theta_s,q_t\}_h=2h(e^{i\theta_s})q_t\delta_{st},
\end{equation}
and then extend it in the canonical fashion:
\begin{equation}\label{E:DefPBh2}
\{f_1,f_2\}_h=
\sum_{s=1}^n 2h(e^{i\theta_s})q_s\left[\frac{\partial f_1}{\partial\theta_s}\frac{\partial f_2}{\partial q_s}
-\frac{\partial f_1}{\partial q_s}\frac{\partial f_2}{\partial\theta_s}\right].
\end{equation}

\begin{prop}
Let $h\,:\, S^1\rightarrow \IR$ be a smooth, nonzero function. Then $\{\cdot,\cdot\}_h$
defined as in \eqref{E:DefPBh2} (or, equivalently, \eqref{E:DefPBh}) is a
Poisson bracket, and, if $h$ is not identically zero on any arc in $S^1$, then $\{\cdot,\cdot\}_h$ is nondegenerate.
Furthermore, any two such brackets are compatible,
in the sense that their sum is again a Poisson bracket.
\end{prop}

\begin{remark}
Note that for $h$ identically equal to 1 we recover the extension
of the Gelfand-Dikij bracket defined in Section~\ref{S:ExtendedPB}. So the proposition
claims that \eqref{E:DefPBh2} defines a family of Poisson brackets compatible
with the GD-bracket.
\end{remark}

\begin{proof}
A simple calculation shows that
$\{\cdot,\cdot\}_h$ obeys the Jacobi identity - note that it is sufficient to check it on
the $\theta_s$'s and $q_t$'s.

If $h$ is not identically zero on any arc in $S^1$, it follows immediately from the definition \eqref{E:DefPBh}
that the $h$-bracket is nondegenerate: indeed,
any smooth, nonconstant function $f$ must depend
nontrivially on at least one of the variables, say $\theta_1$. Then the $h$-bracket of $f$ with
the conjugate variable (in this case $q_1$) will be nonzero:
$$
\{f,q_1\}_h=\frac{\partial f}{\partial\theta_1}\{\theta_1,q_1\}_h
=2h(e^{i\theta_1})q_1\frac{\partial f}{\partial\theta_1}\neq0.
$$
So the only Casimirs are constant functions.

Finally, note that the newly-defined brackets are linear in $h$. In other words, for
any $h_1$ and $h_2$ as above, the sum
$$
\{\cdot,\cdot\}_{h_1}+\{\cdot,\cdot\}_{h_2}\equiv\{\cdot,\cdot\}_{h_1+h_2}
$$
is, by the previous argument, also a Poisson bracket. This is exactly the definition of compatibility.
\end{proof}

Going back to the $\mut_j$ variables, direct calculations show that:
\begin{lemma}
In the notations used above, the $h$-bracket can be written in the $\theta$ and $\mut$ coordinates as:
$$
\{z_s,\mut_t\}_h=2iz_sh(z_s)\mut_t\delta_{st}
$$
and, for $s\neq t$,
$$
\{\mut_s,\mut_t\}_h=i\mut_s\mut_t\bigl(h(z_s)+h(z_t)\bigr)\frac{z_s+z_t}{z_s-z_t}.
$$
\end{lemma}

The analog of the Carath\'eodory function for unnormalized measures is defined, unsurprisingly, by any of the following expressions
\begin{equation}\label{E:DefnCunnorm}
\Ft(z)=\int_{S^1} \frac{\zeta+z}{\zeta-z}\,d\mut(\zeta)=\sum_{j=1}^n \mut_j\frac{z_j+z}{z_j-z}=cF(z),
\end{equation}
where, as before, $c=|\mut|$ is the total weight of the finite measure $\mut$, and $F$ is the usual
Carath\'eodory function associated to the probability measure $\mu=\mut/|\mut|$. Given the approach we take in this paper,
it is natural to try to compute the $h$-bracket of $\Ft$ at two distinct points $z$ and $w$ in the complex plane. The calculation
that will give us these formulae is straightforward enough, with the only caveat that the resulting formula will involve
not only $\Ft$, but also the function
$$
\Ft_h(z)=\sum_{j=1}^n h(z_j)\mut_j\frac{z_j+z}{z_j-z}=\int h(\zeta)\frac{\zeta+z}{\zeta-z}\,d\mut(\zeta).
$$
As before, $\Ft_{h\equiv1}(z)=\Ft(z)$.
Then one has
\begin{equation*}
\begin{aligned}
\{\Ft(z),\Ft(w)\}_h
&=i\frac{w+z}{w-z}\bigl(\Ft(z)-\Ft(w)\bigr)\bigl(\Ft_h(z)-\Ft_h(w)\bigr)\\
&\quad-i\Ft(0)\bigl(\Ft_h(z)-\Ft_h(w)\bigr)-i\Ft_h(0)\bigl(\Ft(z)-\Ft(w)\bigr)\\
&=i(w-z)\sum_{j,k=1}^n \frac{(z_j+z_k)\bigl(h(z_j)+h(z_k)\bigr)(z_jz_k+zw)}{(z_j-z)(z_k-z)(z_j-w)(z_k-w)}\mut_j\mut_k
\end{aligned}
\end{equation*}
While this formula in fairly involved and not very pretty, it simplifies greatly when restricted to the
manifold of probability measures. In order to find this restriction, we need to take the reverse road to that
in the previous section, and hence compute the following bracket:
\begin{equation*}
\begin{aligned}
\{c,\Ft(z)\}_h
&=iz\sum_{j,k=1}^n \frac{(z_j+z_k)\bigl(h(z_j)+h(z_k)\bigr)}{(z_j-z)(z_k-z)}\mut_j\mut_k\\
&=i\bigl[\Ft_h(z)\Ft(z)-\Ft_h(0)\Ft(0)\bigr]
\end{aligned}
\end{equation*}
\begin{equation*}
\{c,F(z)\}_h=ic\bigl[F_h(z)F(z)-F_h(0)\bigr]
\end{equation*}
(where $\Ft_h(z)=cF_h(z)$, and recall $F(0)=1$).
Finally, we obtain the following:
\begin{theorem}
The restrictions of the $h$-brackets to the manifold of probability measures supported at $n$ points
on the unit circle are given by
\begin{equation}\label{E:PBhCara}
\{F(z),F(w)\}_h=i\bigl(F_h(z)-F_h(w)\bigr)\Bigl[\tfrac{w+z}{w-z}\bigl(F(z)-F(w)\bigr)+F(z)F(w)-1\Bigr].
\end{equation}
which defines, for $h$ smooth and real valued, a family of compatible Poisson brackets, that forms
a multi-Hamiltonian structure for the defocusing Ablowitz-Ladik bracket.

In particular, this implies
\begin{equation}\label{E:12}
\{z_s,z_t\}_h=0,\quad\{z_s,\mu_t\}_h=2iz_sh(z_s)\mu_t(\delta_{st}-\mu_s)
\end{equation}
and, for $s\neq t$,
\begin{equation}\label{E:13}
\begin{aligned}
\{\mu_s,\mu_t\}_h=i\mu_s\mu_t\sum_{k\neq s,t} \mu_k
&\Bigl[\frac{\bigl(h(z_s)+h(z_t)\bigr)(z_s+z_t)}{z_s-z_t}
+\frac{\bigl(h(z_t)+h(z_k)\bigr)(z_t+z_k)}{z_t-z_k}\\
&\qquad+\frac{\bigl(h(z_k)+h(z_s)\bigr)(z_k+z_s)}{z_k-z_s}\Bigr]
\end{aligned}
\end{equation}
\end{theorem}

\begin{proof}
The equation \eqref{E:PBhCara} follows directly from the previous formulae and from
$$
\{\Ft(z),\Ft(w)\}_h=c^2\{F(z),F(w)\}_h+cF(z)\{c,F(w)\}_h+cF(w)\{F(z),c\}_h.
$$
The formulae for the brackets of the $z_s$'s and $\mu_t$'s follow from \eqref{E:PBhCara}
by using the residue theorem, exactly as in the proof of Theorem~\ref{T:bracket}.
\end{proof}

Finally, we close this section by identifying Hamiltonians for Ablowitz-Ladik flows in
the $h$-brackets.
\begin{prop}\label{P:5.1}
Let $g$ be a polynomial, and consider the Hamiltonian on $M_n(\IC)$ defined by
$\phi(X)=\Im\Tr(g(X))$. Then the evolution of the spectral measure $\mu=\sum_{j=1}^n \mu_j\delta_{z_j}$, $z_j=e^{i\theta_j}$,
associated to a CMV matrix $\C$ under this Hamiltonian
in the $h$-bracket is given by
\begin{equation}\label{E:17}
\begin{cases}
  \dot z_j &=\{\phi,z_j\}_h=0\\
  \dot\mu_j&=\{\phi,\mu_j\}_h=\mu_j\Bigl[G(z_j)h(z_j)
             -\sum_{l=1}^n G(z_l)h(z_l)\mu_l\Bigr],
\end{cases}
\end{equation}
where $G(z)=2\Re(zg'(z))$. Equivalently,
\begin{equation}\label{E:18}
 \mu(t)=\frac{e^{G(z)h(z)t}\mu(t=0)}{|e^{G(z)h(z)t}\mu(t=0)|}.
\end{equation}
\end{prop}

\begin{proof}
Note that we can rewrite $\phi=\Im\sum_{k=1}^n g(z_k)$.
Then the first formula in \eqref{E:17} follows immediately from \eqref{E:12}, while
\begin{equation*}
 \begin{aligned}
  \dot\mu_j&=\{\phi,\mu_j\}_h=\sum_{k=1}^n \frac{\partial \phi}{\partial \theta_k}\{\theta_k,\mu_j\}_h\\
 \end{aligned}
\end{equation*}
But
$$
\frac{\partial \phi}{\partial \theta_k}=\Im\bigl(ie^{i\theta_k}g'(e^{i\theta_k})\bigr)
=\tfrac12G(e^{i\theta_k}),
$$
and hence
\begin{equation*}
 \begin{aligned}
  \dot\mu_j&=\sum_{k=1}^n G(e^{i\theta_k})h(e^{i\theta_k})\mu_j(\delta_{jk}-\mu_k)\\
           &=\mu_j\Bigl[G(e^{i\theta_j})h(e^{i\theta_j})
             -\sum_{l=1}^n G(e^{i\theta_l})h(e^{i\theta_l})\mu_l\Bigr],
 \end{aligned}
\end{equation*}
as in \eqref{E:17}. Finally, \eqref{E:18} follows from \eqref{E:17} by integration.
\end{proof}

Recall that the Gelfand-Dikij bracket, which is the Poisson bracket associated to the Ablowitz-Ladik equation,
corresponds to $h\equiv1$, while the Hamiltonians which generate the flows in the AL hierarchy are exactly
of the form $\phi(\C)=\Im\Tr (g(\C))$ for a polynomial $g$. So we find that the following holds:
\begin{coro}\label{C:5.1}
If $h$ is a trigonometric polynomial, then the flow generated in the $h$-bracket by
the Hamiltonian $\phi(\C)=\Im\Tr (g(\C))$, where $g$ is a polynomial, is one of Ablowitz-Ladik flows.
\end{coro}

\begin{proof}
Since both $h$ and $G$ are trigonometric polynomials, their product will have the form
$$
h(z)G(z)=\sum_{j=-d}^d c_j z^j,\qquad
\text{where}\quad c_{-j}=\bar c_j.
$$
Then for $z\in S^1$
$$
h(z)G(z)=2\Re \bigl(\sum_{j=1}^d c_j z^j \bigr) +c_0=2\Re(z\tilde g'(z))+c_0=\tilde G(z)+c_0,
$$
where we can choose $\tilde g(z)=\sum_{j=1}^d \frac{c_j}{j} z^j$ (unique up to an additive constant).
If we set $\tilde \phi(\IC)=\Im\Tr(\tilde g(\IC))$, then we see that
$$
\{\phi,\mu_j\}_h-\{\tilde\phi,\mu_j\}_{1}=\mu_j\bigl[c_0-\sum_{k=1}^n c_0\mu_k\bigr]=0,
$$
since $\mu$ is a probability measure. This is exactly what we claimed.
\end{proof}

We close this section by noting an immediate consequence of Corollary~\ref{C:5.1}. Consider
two trigonometric polynomials, $h_j$, $j=1,2$, and two polynomials $g_j$ from which we construct
Hamiltonians as before: $\phi_j(\C)=\Im\Tr(g_j(\C))$. Then $\phi_1$ generates the same flow in
$\{\cdot,\cdot\}_{h_1}$ as $\phi_2$ does in $\{\cdot,\cdot\}_{h_2}$ iff
\begin{equation}\label{E:identflows}
h_1G_1-h_2G_2=\text{const.}\in\IR,
\end{equation}
where, as before, $G_j(z)=2\Re(zg_j'(z))$ for $j=1,2$.

\section{The connection to Schur flows and the Toda lattice}\label{S:6}

Let us now consider the case where the measure $d\mu$ is symmetric
with respect to complex conjugation, or what is equivalent, where
all the Verblunsky parameters are real. In this case, there are
an even number of eigenvalues, $z_1,...,z_n$, $n=2N$, with the extra symmetry
\begin{equation}\label{E:15}
z_{j+N}=\bar z_j,\quad\mu_{j+N}=\mu_j\quad \text{for} \,1\leq j\leq N.
\end{equation}
For simplicity of the notation, we further assume that
$z_1,...,z_N$ are the eigenvalues on the upper half of the unit circle.
It is a famous observation of
Szeg\H{o} (see \cite[\S11.5]{Szego}) that the polynomials orthogonal
with respect to this measure are intimately related to the
polynomials orthogonal with respect to the measure $d\nu$ on
$[-2,2]$ defined by
\begin{equation}\label{nuDefn}
  \int_{S^1} f(z+z^{-1}) \, d\mu(z)  =  \int_{-2}^2 f(x) \,d\nu(x).
\end{equation}
The recurrence coefficients for these systems of orthogonal polynomials are related by the Geronimus
relations:
\begin{equation}\label{AB:Geron}
\begin{cases}
b_{k+1}   &= (1-\alpha_{2k-1})\alpha_{2k} -
(1+\alpha_{2k-1})\alpha_{2k-2}           \\
a_{k+1}   &= \big\{
(1-\alpha_{2k-1})(1-\alpha_{2k}^2)(1+\alpha_{2k+1}) \big\}^{1/2}.
\end{cases}
\end{equation}
It is an easy observation (see, for example, \cite{N2}) that the second flow in the Ablowitz-Ladik hierarchy,
which is generated by $\Im\Tr\C$, will preserve the property of all the Verblunsky coefficients being in $(-1,1)$.
Hence it makes sense to ask what is the flow it induces via the Geronimus relations to the $a$'s and $b$'s. A direct
calculation shows that the answer is exactly the Toda flow! It is immediately
clear from \eqref{E:ALhimag} that the submanifold of real Verblunsky coefficients, which we will denote by $M$, is in fact
stable under any of
the flows generated in the usual AL (or GD) bracket by the Hamiltonians $\Im K_k=\frac1k \Im\Tr (\C^k)$, $k\geq1$.
We will call these {\em the Schur flows} ( cf. \cite{FG1, Gol}.
But neither the Hamiltonian in question, nor the usual Ablowitz-Ladik
Poisson bracket have meaningful restrictions to this submanifold, nor is it possible to find the image of the flows generated
by $\Im K_k$ under the Geronimus relations by straightforward calculations.
In this section we investigate the newly defined $h$-brackets from these points of view.

\begin{prop}\label{P:6.1}
The $h$-bracket $\{\cdot,\cdot\}_h$ has a restriction to the submanifold $M$ of probability measures
on the unit circle which are symmetric
with respect to complex conjugation iff $h(\bar z)=-h(z)$ for $z\in S^1$.
\end{prop}

\begin{proof}
To prove this statement, consider a set of functions on the manifold of measures supported at $n=2N$ points defined by
\begin{equation}
\la_s=z_s+\bar z_s\quad\text{and}\quad \nu_s=2\mu_s.
\end{equation}
In fact, these functions with $1\leq s\leq N$ form a set of coordinates on the submanifold of probability measures on the unit
circle which are symmetric with respect to complex conjugation, and
\begin{equation}\label{E:14}
\la_s=z_s+z_{s+N},\qquad \nu_s=\mu_s+\mu_{s+N}
\end{equation}
for $1\leq s\leq N$. Furthermore, the measure $d\nu$ on $[-2,2]$ defined above is exactly
$d\nu=\sum_{s=1}^N \nu_s \delta_{\la_s}.$ Direct calculation using \eqref{E:12} shows that
$$
\{\la_s,\la_t\}_h=0,\quad\text{for every}\quad 1\leq s,t\leq N.
$$
The other two types of brackets are more complicated. Decompose a general function $h$ as $h=h_++h_-$,
where $h_+(z)=h_+(\bar z)$ and $h_-(z)=-h_-(\bar z)$ for $z\in S^1$. This is equivalent to setting
$2h_+(z)=h(z)+h(\bar z)$ and $2h_-(z)=h(z)-h(\bar z)$.
Then consider $1\leq s,t\leq N$, and use \eqref{E:14}, \eqref{E:15}, and \eqref{E:12} with $s,s+N,t,$ and $t+N$, respectively:
\begin{equation*}
\begin{aligned}
\{\la_s,\nu_t\}_h
&=2iz_sh(z_s)\mu_t(\delta_{st}-\mu_s)+2iz_sh(z_s)\mu_{t+N}(-\mu_s)\\
&\quad+2iz_{s+N}h(z_{s+N})\mu_t(-\mu_{s+N})+2iz_{s+N}h(z_{s+N})\mu_{t+N}(\delta_{st}-\mu_{s+N})\\
&=h_+(z_s)\bigl[2i(z_s+\bar z_s)\mu_t(\delta_{st}-\mu_s)-2i(z_s+\bar z_s)\mu_s\mu_t\bigr]\\
&\quad+h_-(z_s)\bigl[2i(z_s-\bar z_s)\mu_t(\delta_{st}-\mu_s)-2i(z_s-\bar z_s)\mu_s\mu_t\bigr]\\
&=2i(z_s+\bar z_s)h_+(z_s)\mu_t(\delta_{st}-2\mu_s)\\
&\quad+2i(z_s-\bar z_s)h_-(z_s)\mu_t(\delta_{st}-2\mu_s)
\end{aligned}
\end{equation*}
The last expression is real-valued iff $h_+\equiv 0$, or, equivalently, $h(\bar z)=-h(z)$. In this case,
we get
$$
\{\la_s,\nu_t\}_h=2i(z_s-\bar z_s)h_-(z_s)\mu_t(\delta_{st}-2\mu_s),
$$
where the right-hand side is real for $z_s\in S^1$, and invariant under the mapping taking a
probability measure to its complex conjugate.

Finally, we need to deal with the $h$-bracket of the $\nu$'s. Proceeding as above, we get:
\begin{equation*}
\begin{aligned}
\{\nu_s,\nu_t\}_h=\{\mu_s,\mu_t\}_h+\{\mu_{s+N},\mu_{t+N}\}_h+\{\mu_{s+N},\mu_t\}_h+\{\mu_s,\mu_{t+N}\}_h
\end{aligned}
\end{equation*}
Group the first two terms on the right-hand side to get
\begin{align*}
&i\mu_s\mu_t\sum_{k\neq s,t} \mu_k
\Bigl[\frac{\bigl(h(z_s)+h(z_t)\bigr)(z_s+z_t)}{z_s-z_t}
+\frac{\bigl(h(z_t)+h(z_k)\bigr)(z_t+z_k)}{z_t-z_k}
+\frac{\bigl(h(z_k)+h(z_s)\bigr)(z_k+z_s)}{z_k-z_s}\Bigr]\\
&\begin{aligned}+i\mu_{s+N}\mu_{t+N}\sum_{l\neq s+N,t+N} \mu_l
\Bigl[&\frac{\bigl(h(z_{s+N})+h(z_{t+N})\bigr)(z_{s+N}+z_{t+N})}{z_{s+N}-z_{t+N}}
+\frac{\bigl(h(z_{t+N})+h(z_l)\bigr)(z_{t+N}+z_l)}{z_{t+N}-z_l}\\
&\qquad+\frac{\bigl(h(z_l)+h(z_{s+N})\bigr)(z_l+z_{s+N})}{z_l-z_{s+N}}\Bigr]
\end{aligned}\\
&\begin{aligned}=i\mu_s\mu_t\sum_{k\neq s,t} \mu_k\Bigl[&\frac{\bigl(h(z_s)+h(z_t)\bigr)(z_s+z_t)}{z_s-z_t}+
\frac{\bigl(h(z_{s+N})+h(z_{t+N})\bigr)(z_{s+N}+z_{t+N})}{z_{s+N}-z_{t+N}}\\ &+\text{cyclic permutations}\Bigr]
\end{aligned}
\end{align*}
To obtain this last identity, set $l=k+N$ in the second sum on the left-hand side, and use formula \eqref{E:13} as well as the symmetry conditions
\eqref{E:15}. Note that we think of the indices as periodic, with period $2N$. The main observation at this point is that, for any two indices
$1\leq l,r\leq N$, we have
\begin{equation*}
\begin{aligned}
&\frac{\bigl(h(z_l)+h(z_r)\bigr)(z_l+z_r)}{z_l-z_r}+\frac{\bigl(h(z_{l+N})+h(z_{r+N})\bigr)(z_{l+N}+z_{r+N})}{z_{l+N}-z_{r+N}}\\
&=(h_+(z_l)+h_+(z_r))\Bigl[\frac{z_l+z_r}{z_l-z_r}+\frac{\bar z_l+\bar z_r}{\bar z_l-\bar z_r}\Bigr]
+(h_-(z_l)+h_-(z_r))\Bigl[\frac{z_l+z_r}{z_l-z_r}-\frac{\bar z_l+\bar z_r}{\bar z_l-\bar z_r}\Bigr]\\
\end{aligned}
\end{equation*}
But it is a simple observation that, for $z,w\in S^1$,
$$
\frac{\bar z+\bar w}{\bar z-\bar w}=-\frac{z+w}{z-w},
$$
and hence the expression above equals
$$
2(h_-(z_l)+h_-(z_r))\frac{z_l+z_r}{z_l-z_r}.
$$
In particular, this means that we can work our way backwards to the original expressions for the four $h$-brackets, only
with $h$ replaced by $h_-$. Note that, for any $z\neq w\in S^1$, $i\frac{z+w}{z-w}\in \IR$ and
$$
\frac{\bigl(h_-(z)+h_-(w)\bigr)(z+w)}{z-w}=\frac{\bigl(h_-(\bar z)+h_-(\bar w)\bigr)(\bar z+\bar w)}{\bar z-\bar w}.
$$
But this immediately implies that, even though the bracket $\{\nu_s,\nu_t\}_h$ is still the sum of four complicated formulae,
it is real-valued and invariant under the mapping taking a
probability measure to its complex conjugate, which completes the proof.
\end{proof}

Now consider a function $h$ obeying
$$
h(e^{-i\theta})=-h(e^{i\theta}),
$$
and restrict the bracket $\{\cdot,\cdot\}_h$ to the subspace of
real Verblunsky coefficients.
We want to write this restriction as a combination
of the (compatible) Poisson brackets for the Toda lattice found in \cite{FG2}. To avoid
confusion, we will go back to the notation from Section~\ref{S:3}. Hence we will start denoting
the $h$-brackets by $\{\cdot,\cdot\}^{(1)}_h$, the superscript denoting the fact that this is a Poisson bracket
compatible with the Gelfand-Dikij bracket. By contrast, we will later on be interested in some
of the Poisson brackets compatible with the Lie-Poisson bracket, which were originally introduced
in \cite{FG2}, and which we will denote here by $\{\cdot,\cdot\}^{(0)}_H$, for some function $H$.

In order to achieve this, we must first relate the Carath\'eodory function of a measure $d\mu$ on the circle
which is invariant under complex conjugation to the $m$-function of the associated measure $d\nu$ on $[-2,2]$.
A simple calculation (see, for example, \cite{Simon1}) show that, for $z\in\IC\setminus\{0\}$, we have
\begin{equation}\label{E:CaraM}
F(z)=-F\bigl(\tfrac1z\bigr)=\bigl(z-\tfrac1z\bigr)\cdot m\bigl(z+\tfrac1z\bigr),
\end{equation}
and
\begin{equation}\label{E:CaraMH}
F_h(z)=F_h\bigl(\tfrac1z\bigr)=im_H\bigl(z+\tfrac1z\bigr),
\end{equation}
where
\begin{equation}\label{E:DefH}
H(2\cos\theta)=2\sin(\theta) h(e^{i\theta}),
\end{equation}
$h$ obeys $h(e^{-i\theta})=-h(e^{i\theta})$ as above, and $m_H$ is defined by
\begin{equation}\label{E:DefmH}
m_H(\lambda)=\int_{-2}^2 H(t)\frac{1}{t-\lambda}d\nu(t).
\end{equation}
Note that, for $z\in S^1$, $H$ is defined by
$$
H(z+\bar z)=(z-\bar z)h(z),
$$
which is exactly the type of combination that has already appeared in the proof of Proposition~\ref{P:6.1}.

We wish to find the $h$-bracket of the $m$ function and compare it to the ones in \cite{FG2}. To do so,
we insert \eqref{E:CaraM} into \eqref{E:PBhCara}, but
we must not lose sight of the symmetry inherent to the situation. In this case, we use the fact that
\begin{equation}\label{E:PBCaraM}
\begin{aligned}
4\bigl(z-\tfrac1z\bigr)\bigl(w-\tfrac1w\bigr)
\bigl\{m\bigl(z+\tfrac1z\bigr),m\bigl(w+\tfrac1w\bigr)\bigr\}^{(1)}_h
&=\bigl\{F(z)-F\bigl(\tfrac1z\bigr),F(w)-F\bigl(\tfrac1w\bigr)\bigr\}^{(1)}_h
\end{aligned}
\end{equation}
Using \eqref{E:CaraM}, we get that
\begin{equation*}\label{E:PBCaraM2}
\begin{aligned}
\bigl\{F(z)-F\bigl(\tfrac1z\bigr),F(w)-F\bigl(\tfrac1w\bigr)\bigr\}^{(1)}_h
&=\bigl\{F(z),F(w)\bigr\}^{(1)}_h-\bigl\{F\bigl(\tfrac1z\bigr),F(w)\bigr\}^{(1)}_h\\
&\quad-\bigl\{F(z),F\bigl(\tfrac1w\bigr)\bigr\}^{(1)}_h+\bigl\{F\bigl(\tfrac1z\bigr),F\bigl(\tfrac1w\bigr)\bigr\}^{(1)}_h\\
&=(F_h(z)-F_h(w))\cdot \bigl(B(z,w)-4F(z)F(w)\bigr)
\end{aligned}
\end{equation*}
where
\begin{equation*}
\begin{aligned}
B(z,w)
&=-\frac{w+z}{w-z}\bigl(F(z)-F(w)\bigr)+\frac{w+\tfrac1z}{w-\tfrac1z}\bigl(-F(z)-F(w)\bigr)\\
&\quad +\frac{\tfrac1w+z}{\tfrac1w-z}\bigl(F(z)+F(w)\bigr)-\frac{\tfrac1w+\tfrac1z}{\tfrac1w-\tfrac1z}\bigl(-F(z)+F(w)\bigr)\\
&=4z\frac{1-w^2}{(z-w)(1-zw)}F(z)-4w\frac{1-z^2}{(z-w)(1-zw)}F(w)\\
&=4\bigl(z-\tfrac1z\bigr)\bigl(w-\tfrac1w\bigr)\frac{m\bigl(z+\tfrac1z\bigr)-m\bigl(w+\tfrac1w\bigr)}{\bigl(z+\tfrac1z\bigr)
-\bigl(w+\tfrac1w\bigr)}
\end{aligned}
\end{equation*}

Putting it all together and replacing the combination $z+\frac1z$ by a general $\lambda\in\IC\setminus \text{spec}(d\nu)$, we get that
\begin{equation*}
\{m(\lambda),m(\xi)\}^{(1)}_h=(m_H(\lambda)-m_H(\xi))\cdot\Bigl[\frac{m(\lambda)-m(\xi)}{\lambda-\xi}-m(\lambda)m(\xi)\Bigr]
\end{equation*}

Let us recall that, for $k\geq0$, Faybusovich and Gekhtman defined
a Poisson bracket on the space of measures supported on $[-2,2]$. This $k$-bracket can be written down in terms of
various coordinates, but here we concentrate on its expression in terms of the associated $m$-functions:
\begin{equation}\label{E:kPB}
\{m(\lambda),m(\xi)\}_k
=\bigl((\lambda^km(\lambda))_--(\xi^km(\xi))_-)\bigr)\cdot
\Bigl[\frac{m(\lambda)-m(\xi)}{\lambda-\xi}-m(\lambda)m(\xi)\Bigr]\footnote{We use here
the notation of  \cite{FG2} for the compatible brackets.},
\end{equation}
where, for a meromorphic function $r\,:\,\IC\rightarrow\IC$ with $r(\lambda)=\sum_{l=-\infty}^N c_l\lambda^l$, $N<\infty$,
we set
$$
(r(\lambda))_+=\sum_{l=0}^N c_l \lambda^l \quad \text{and} \quad (r(\lambda))_-=r(\lambda)-(r(\lambda))_+=
\sum_{l=-\infty}^{-1} c_l\lambda^l.
$$
Note that for any $k\geq0$ we have
$$
\Bigl(\frac{t^k-\lambda^k}{t-\lambda}\Bigr)_-=\Bigl(\sum_{j=0}^{k-1}t^j\lambda^{k-j-1}\Bigr)_-=0,
$$
and hence
$$
\Bigl(\frac{\lambda^k}{t-\lambda}\Bigr)_-=\Bigl(\frac{t^k}{t-\lambda}\Bigr)_-=\frac{t^k}{t-\lambda}.
$$
So by integrating, we get that
$$
\bigl(H(\lambda)m(\lambda)\bigr)_-=m_H(\lambda),
$$
where $H$ is a polynomial and $m_H$ is defined by \eqref{E:DefmH}. In other words, we have proved the following
\begin{prop}\label{P:6.2}
If $h$ is a smooth, real-valued function on the unit circle $S^1$ such that $h(\bar z)=-h(z)$, then the
restriction of the bracket $\{\cdot,\cdot\}^{(1)}_h$ to the submanifold of probability measures
invariant under complex conjugation is given, for $\la\neq\xi \in \IC\setminus \IR$, by
\begin{equation}\label{E:HPB}
\{m(\lambda),m(\xi)\}^{(1)}_h=(m_H(\lambda)-m_H(\xi))\cdot\Bigl[\frac{m(\lambda)-m(\xi)}{\lambda-\xi}-m(\lambda)m(\xi)\Bigr],
\end{equation}
where $H$, $m$ and $m_H$ are defined as above.

The right-hand side defines a Poisson bracket
$\{\cdot,\cdot\}^{(0)}_H$ on the manifold of probability measures
supported at $N$ points on $[-2,2]$ which is compatible with the
Toda lattice (i.e. restriction of the Lie-Poisson) bracket.
Furthermore, if $h$ is a trigonometric polynomial, then $H$ is a
polynomial and $\{\cdot,\cdot\}^{(0)}_H$ is a linear combination of
the compatible brackets \eqref{E:kPB} of \cite{FG2}.
\end{prop}

While we could try to write down the general formula for these brackets,
we will limit ourselves to investigating the simplest case, which is
$$
h(e^{i\theta})=2\sin(\theta).
$$
Then we get that $H(2\cos(\theta))=4\sin(\theta)^2$, or, equivalently,
$$
H(t)=4-t^2, \qquad\text{for}\quad t\in [-2,2].
$$
Therefore we find that the restriction of the Poisson bracket $\{\cdot,\cdot\}^{(1)}_{2\sin}$ to
the space of real Verblunsky coefficients coincides, under the Geronimus relations,
to the Poisson bracket $\{\cdot,\cdot\}^{(0)}_H=4\{\cdot,\cdot\}_0-\{\cdot,\cdot\}_2$.

We close this paper by identifying  the Hamiltonians defining certain Ablowitz-Ladik flows with
Toda hierarchy Hamiltonians. Recall that the usual AL bracket is $\{\cdot,\cdot\}_{h_1}$ for $h_1\equiv1$,
and let $h_2(\bar z)=-h_2(z)$ for $z\in S^1$, as in Proposition~\ref{P:6.1}. Consider a polynomial with
real coefficients, $g_1$, and note that the Hamiltonian $\phi_1(\C)=\Im\Tr g_1(\C)$ is a linear combination,
with real coefficients, of the flows $\Im K_k$ and hence it generates a flow under which $M$ is stable. If
$g_2$ is another polynomial and $\phi_2(\C)=\Im\Tr g_2(\C)$, then from \eqref{E:identflows} we get that if
\begin{equation}\label{E:6.1}
G_1(z)-h_2(z)G_2(z)\equiv \text{const.}\in\IR,
\end{equation}
then the two flows coincide:
\begin{equation*}
\{\phi_1,\cdot\}_{_{GD}}\equiv \{\phi_2,\cdot\}_{h_2}\,.
\end{equation*}
Here, as in Section~\ref{S:CompatiblePB}, $G_j(z)=2\Re(zg_j'(z))$. Let us make a few remarks on \eqref{E:6.1}:\\
$\bullet$ Since $g_1(z)=\sum_{j=0}^d c_j z^j$ with $c_j\in\IR$, we get $G_1(z)=\sum_{j=1}^d jc_j(z^j+\bar z^j)$ and so for $z\in S^1$,
$G_1(\bar z)=G_1(z)$. Since we imposed $h_2(\bar z)=-h_2(z)$, we must have that $G_2(\bar z)=-G_2(z)$, or, if we work as
in the proof of Corollary~\ref{C:5.1}, the coefficients of $g_2$ are purely imaginary, up
to an additive constant which we will ignore since it does not influence the flow. So the first observation
is that
\begin{equation}
\phi_2(\C)=\Re\Tr (-i g_2(\C))
\end{equation}
is a linear combination with real coefficients of the Hamiltonians $\Re K_k$, and hence has a nontrivial
restriction to the submanifold $M$.\\
$\bullet$ A straightforward count of the parameters in \eqref{E:6.1}
shows that it is not true that given $g_1$, we can always find $h_2$
and $G_2$ with the required properties and which satisfy
\eqref{E:6.1}. Indeed, without loss of generality we may assume that
$g_1$, $h_2$ and $G_2$ are monic. Then the right-hand side of
\eqref{E:6.1} is determined by $\deg(h_2)-1+\deg(G_2)-1=\deg(G_1)-2$
parameters, while the left-hand side imposes $\deg(G_1)-1$
conditions. In other words, we need an extra degree of freedom.
Without going into too many details, let us mention that one way to
deal with this problem is to allow the Schur flow in question to be
modified by a constant multiple of the first Schur flow: for any
$g_1$ monic of degree at least 2 and with real coefficients, there
exist a real constant $c$, and monic $h_2$ and $g_2$ as above so
that $\tilde G_1(z)=2\Re(zg_1'(z)-cz)$, $h_2$ and $G_2$ obey
\eqref{E:6.1}. In particular, the restriction to the submanifold $M$
of a Schur flow whose $G_1$ obeys \eqref{E:6.1} for some appropriate
$h_2$ and $G_2$ is a Hamiltonian flow in the Poisson bracket
$\{\cdot,\cdot\}\upharpoonright_M$.\footnote{Since $h_2(\bar
z)=-h_2(z)$,
we know from Proposition~\ref{P:6.1} that $M$ is a Poisson submanifold in this Poisson structure.}\\
$\bullet$ Finally, we must understand the image through the Geronimus relations of the restriction to $M$
of Hamiltonians given by $\Re(g(\C))$, with  $g$ polynomial with real coefficients. But this can be obtained immediately
from Proposition~B.3 of Killip and Nenciu \cite{KN1}, which shows that, if $\al_j\in (-1,1)$
for all $j$, then $\C+\C^T$ is unitarily equivalent to a direct sum of two Jacobi matrices, $J$
and $\tilde J$, where entries of $J$ are related to the Verblunsky coefficients defining $\C$ by the Geronimus relations.
Furthermore, the spectral measure for $J$ w.r.t. the vector $e_1=[1\,0\cdots]^T\in\IR^n$
is $d\nu$, while the spectral measure of $\tilde J$ w.r.t. $e_1$ is $d\tilde\nu(x)=\frac1{2(1-\alpha_0^2)(1-\alpha_1)}(4-x^2)\,d\nu(x)$.
In particular, we see that the two measures $\nu$ and $\tilde\nu$ have the same support, or, equivalently,
$J$ and $\tilde J$ have the same eigenvalues.
For example,
\begin{equation*}
\Re\Tr\C=\tfrac12\Tr(\C+\C^T)=\tfrac12\Tr(J\oplus\tilde J)=\Tr(J).
\end{equation*}
A slightly more careful analysis will show that for any $k\geq1$ there exists a monic polynomial $g_k$, with real
coefficients, so that
\begin{equation*}
\Re\Tr (\C^k)\upharpoonright_M=\Tr(g_k(J)).
\end{equation*}
But the Hamiltonians on the right-hand side are exactly Toda hierarchy Hamiltonians.



\begin{thebibliography}{10}

\bibitem[AblLad1]{AL1} M.~J.~Ablowitz, J.~F.~Ladik, Nonlinear
differential-difference equations. \textit{J. Math. Phys.} {\bf 16}
(1975), 598--603.

\bibitem[AblLad2]{AL2} M.~J.~Ablowitz, J.~F.~Ladik, Nonlinear
differential-difference equations and Fourier analysis. \textit{J.
Math. Phys.} {\bf 17} (1976), 1011--1018.

\bibitem[AblPriTru]{APT} M.~J.~Ablowitz, B.~Prinari, A.~D.~Trubach, \textit{Discrete and Continuous Nonlinear Schr\"odinger
Systems}. London Mathematical Society Lecture Note Series, Vol. 302,
Cambridge University Press, Cambridge, 2004.

\bibitem[Adl]{Adler} M.~Adler,
On a trace functional for formal pseudo differential operators and
the symplectic structure of
    the Korteweg-de\thinspace Vries type equations.
\textit{Invent. Math.} \textbf{50} (1979), 219--248.

\bibitem[BloGek]{BG} A.~M.~Bloch and M.~Gekhtman, Lie algebraic aspects of the finite nonperiodic Toda flows.
\textit{J. Comp. $\&$ Appl. Math.} \textbf{202} (2007), 3--25.

\bibitem[CanMorVel1]{CMV} M.~J.~Cantero, L.~Moral, L.~Vel\'azquez,
Five-diagonal matrices and zeros of orthogonal polynomials on the
unit circle. \textit{Linear Algebra Appl.} {\bf 362} (2003), 29--56.

\bibitem[CanMorVel2]{CMV2} M.~J.~Cantero, L.~Moral, L.~Vel\'azquez,
Minimal representations of unitary operators and orthogonal
polynomials on the unit circle, preprint, \texttt{arXiv:math.CA/0405246}.

\bibitem[CanSim]{CanSim} M.~J.~Cantero, B.~Simon, Poisson brackets of orthogonal polynomials, preprint,
\texttt{arXiv:math/0610989}.

\bibitem[Com]{Common}
A.~K.~Common,
A solution of the initial value problem for half-infinite integrable lattice systems.
\textit{Inverse Problems} {\bf  8} (1992) 393--408.

\bibitem[Dei]{Deift} P.~Deift, Integrable Hamiltonian Systems. \textit{Dynamical systems and probabilistic
methods in partial differential equations (Berkeley, CA, 1994)},
103--138, Lectures in Appl. Math., {\bf 31}, Amer. Math. Soc.,
Providence, RI, 1996.

\bibitem[DLNT1]{DLNT_CPAM86} P.~Deift, L.~C.~Li, T.~Nanda, and C.~Tomei,
The Toda flow on a generic orbit is integrable. \textit{Comm. Pure
Appl. Math.} \textbf{39} (1986), 183--232.

\bibitem[FayGek1]{FG1} L.~Faybusovich and M.~Gekhtman, On Schur flows.
\textit{J. Phys. A: Math. Gen.} {\bf 32} (1999), 4671--4680.

\bibitem[FayGek2]{FG2} L.~Faybusovich and M.~Gekhtman, Poisson brackets on rational functions
and multi-Hamiltonian structure for integrable lattices.
\textit{Phys. Lett. A} {\bf 272} (2000), no. 4, 236--244.

\bibitem[Fla1]{Fla1} H.~Flaschka,
The Toda lattice. I. Existence of integrals. \textit{Phys. Rev. B}
\textbf{9} (1974), 1924--1925.

\bibitem[Fla2]{Fla} H.~Flaschka, Discrete and periodic illustrations of some aspects of the inverse
method, \textit{Dynamical Systems, Theory and Applications} (Rencontres, Battelle Res. Inst., Seattle, Wash., 1974), pp.~441--466,
Lecture Notes in Phys., {\bf 38}, Springer, Berlin, 1975.

\bibitem[Gek]{G} M.~Gekhtman, Non-Abelian nonlinear lattice equations on finite interval,
\textit{J. Phys. A: Math. Gen.} {\bf 26} (1993), 6303--6317.

\bibitem[Gol]{Gol} L.~B.~Golinskii, Schur flows and orthogonal polynomials on the unit circle.
\textit{Mat. Sb.} {\bf  197}  (2006)  41--62.

\bibitem[GelDik]{GD} I.~M.~Gelfand and L.~A.~Dikij,
A family of Hamiltonian structures related to nonlinear integrable
differential equations. \textit{Izrail~M.~Gelfand Collected Papers,
vol. 1}, Springer-Verlag 1987, 625--646.

\bibitem[KilNen1]{KN1} R.~Killip and I.~Nenciu, Matrix models for circular ensembles,
\textit{Int. Math. Res. Not.} \textbf{50} (2004), 2665-2701.

\bibitem[KilNen2]{KN} R.~Killip, I.~Nenciu, CMV: the unitary analogue of
Jacobi matrices. Preprint, to appear in \textit{Comm. Pure Appl. Math.}

\bibitem[Kos]{Kostant} B.~Kostant, The solution to a generalized Toda lattice and representation theory.
\textit{Adv. Math.} {\bf 34} (1979) 195--338.

\bibitem[Li]{L} L.-C.~Li, Some remarks on CMV matrices and dressing
orbits, \textit{Int. Math. Res. Not.} {\bf 40} (2005), 2437--2446.

\bibitem[Mos]{Moser} J.~Moser, Finitely many mass points on the line under the influence of an exponential
potential--an integrable system. \textit{ Battelles Rencontres, Springer  Lecture Notes in Phys.} {\bf 38} (1975) 417--497.

\bibitem[Nen1]{N} I.~Nenciu, Lax pairs for the Ablowitz-Ladik system via orthogonal polynomials on the unit
circle, \textit{Int. Math. Res. Not.} {\bf 11} (2005), 647--686.

\bibitem[Nen2]{N2} I.~Nenciu, Lax Pairs for the Ablowitz-Ladik System via Orthogonal Polynomials on the Unit Circle,
Ph.D. Thesis, Caltech, May 2005.

\bibitem[Nen3]{N3} I.~Nenciu, Poisson brackets for orthogonal polynomials on the unit circle,
preprint, \texttt{arXiv:math.CA/0701055}.

\bibitem[OPRS]{IntSysII} M.~A.~Olshanetsky, A.~M.~Perelomov, A.~G.~Reyman and M.~A.~Semenov-Tian-Shansky,
Integrable systems. II.  \textit{Dynamical systems. VII.}
Encycl. Math. Sci. \textbf{16}, 83--259.

\bibitem[Per]{Perel} A.~M.~Perelomov,
\textit{Integrable systems of classical mechanics and Lie algebras.
Vol. I.} Translated from the Russian by A. G. Reyman. Birkh\"auser
Verlag, Basel, 1990.

\bibitem[Sim1]{Simon1} B.~Simon, \textit{Orthogonal Polynomials on the
Unit Circle, Part~1: Classical Theory}, AMS Colloquium Series,
American Mathematical Society, Providence, RI, 2005.

\bibitem[Sim2]{Simon2} B. Simon, \textit{Orthogonal Polynomials on the Unit Circle, Part~2: Spectral Theory},
AMS Colloquium Series, American Mathematical Society, Providence, RI, 2005.

\bibitem[Sem1]{STS1} M.~Semenov-Tian-Shansky, What is a classical r-matrix?
\textit{Funct. Anal. Appl.} {\bf 17} (1983), 259-272

\bibitem[Sem2]{STS2} M.~Semenov-Tian-Shansky, Dressing transformations and
Poisson group actions. \textit{Publ. RIMS, Kyoto University} {\bf
21} (1985), 1237-1260

\bibitem[Sze]{Szego} G.~Szeg\H{o},
\textit{Orthogonal Polynomials.} American Mathematical Society
Colloquium Publications, Vol. XXIII. American Mathematical Society,
Providence, Rhode Island, 1975.

\bibitem[Tsi]{Tsi} A.~V.~Tsiganov, A family of the Poisson brackets compatible with the Sklyanin bracket,
preprint, \texttt{arXiv:nlin/0612025v2}.
\end{thebibliography}
\end{document}